\documentclass[aps,prb,reprint,superscriptaddress,
amsmath,citeautoscript,flushbottom,nobalancelastpage,floatfix]{revtex4-2}

\usepackage{graphicx}
\usepackage[colorlinks=true,linkcolor=black,citecolor=blue,urlcolor=blue]{hyperref}

\newcommand{\vc}[1]{\boldsymbol{#1}}
\newcommand{\pd}{{\phantom{\dagger}}}
\newcommand{\eeQR}{\mathrm{e}^{i\vc Q\cdot\vc R}}

\begin{document}

\title{Magnetic polarons due to spin-length fluctuations 
in $\boldsymbol{d}^\mathbf{4}$ spin-orbit Mott systems}

\author{Jan Revenda}

\affiliation{Department of Condensed Matter Physics, Faculty of Science,
Masaryk University, Kotl\'a\v{r}sk\'a 2, 61137 Brno, Czech Republic}

\author{Krzysztof Wohlfeld}
\affiliation{Institute of Theoretical Physics, Faculty of Physics,
University of Warsaw, Pasteura 5, PL-02093 Warsaw, Poland}
\affiliation{Department of Physics and Quantum Centre of Excellence for Diamond and
Emergent Materials (QuCenDiEM), Indian Institute of Technology Madras,
Chennai 600036, India
}

\author{Ji\v{r}\'{\i} Chaloupka}

\affiliation{Department of Condensed Matter Physics, Faculty of Science,
Masaryk University, Kotl\'a\v{r}sk\'a 2, 61137 Brno, Czech Republic}

\begin{abstract}
Mott insulators based on $4d$ and $5d$ transition-metal ions, where spin-orbit
interaction plays a key role, can exhibit various forms of unusual magnetism.
A particular example is the antiferromagnet \mbox{Ca$_2$RuO$_4$} containing
$d^4$ \mbox{Ru$^{4+}$} ions. Here the spin-orbit interaction stabilizes the
non-magnetic $J=0$ singlet ionic ground state, which gets dynamically 
mixed---via exchange interactions---with low-energy $J=1$ ionic excitations. Thanks
to a sufficient strength of the exchange, these excitations condense and a
long-range order emerges. The resulting ordered moments are soft and prone to
fluctuations of their effective length. The corresponding amplitude mode
appears as a prominent magnetic excitation and complements the conventional
magnons involving rotations of the moments.
Motivated by this peculiar kind of magnetic order and the specific spectrum of
magnetic excitations, we study their influence on the propagation of doped
carriers. To this end, we construct a microscopic model including both $d^4$
and $d^5$ degrees of freedom and address the propagation of an injected
electron by employing self-consistent Born approximation. We find that the
electron shows a combination of both free and a polaronic type of motion,
where the mobile carrier strongly interacts with an accompanying cloud of
magnetic excitations. Remarkably, in the latter case it is the exotic
excitation---the amplitude mode---that is found to dominate over the
contribution of magnons. Our soft-spin situation thus largely contrasts with
spin polarons widely discussed in the context of doped Heisenberg-like magnets
based on rigid spin moments.
\end{abstract}

\date{\today}

\maketitle


\section{Introduction}

Mott insulators with active $3d$ orbitals in the valence shell have
traditionally been described by the so-called Kugel-Khomskii spin-orbital
superexchange models~\cite{Kug82, Tok00}. These Hamiltonians may host spin and
orbitally ordered ground states as well as collective excitations carrying
spin (magnon)~\cite{Mou96, Ole00, Kha03} or orbital (orbiton)~\cite{Woh11,
Sch12, Mar24} quantum numbers. 
While such physics is rich and far from being completely understood in itself,
another important element has been stressed by Jackeli and Khaliullin in
2009~\cite{Jac09}, namely the large spin-orbit coupling (SOC) present in ions
with $4d$ or $5d$ valence orbitals, which is capable to fundamentally change
the low-energy description of the corresponding Mott insulators~\cite{Wit14,
Tak21}.
In their example, a Mott insulator with $t^5_{2g}$ ions on the honeycomb
lattice is, in the limit of dominant SOC, described by an effective $J=\frac12$
pseudospin model hosting frustrated interactions of Kitaev type. Such an
unexpected possibility of a material realization of the highly interesting
Kitaev model~\cite{Kit06} featuring a spin liquid ground state attracted a lot
of attention and opened an entirely new field of ``Kitaev materials'' such as
\mbox{Na$_2$IrO$_3$} or \mbox{$\alpha$-RuCl$_3$} \cite{Win17,Tak19,Mat25}.

Our paper is inspired by another peculiar spin-orbit Mott insulator,
\mbox{Ca$_2$RuO$_4$}, featuring four valence electrons in $t_{2g}$ orbitals of
$4d^4$ ruthenium Ru$^{4+}$ ions forming square-lattice planes. Since its
discovery in 1997 \cite{Nak97} and observation of the metal-insulator
transition (MIT) at $T_S=357\:\mathrm{K}$~\cite{Bra98, Ale99, Gor10} as well
as onset of the AF order at $T_\mathrm{N}=110\:\mathrm{K}$~\cite{Cao97}, this
ruthenate is subject of ongoing extensive research. For instance, very
recently several phenomena related to MIT due to surface doping~\cite{Hor23},
uniaxial strain~\cite{Ric18}, or d.c. current~\cite{Sue24} were discussed.
Here we choose \mbox{Ca$_2$RuO$_4$} because the effectively moderate value of
its SOC leads to a rich physics, perhaps as fascinating as that of the
above-mentioned Kitaev materials. In fact, the strongly competing spin-orbital
superexchange and SOC in this compound has been intensely debated in the
literature~\cite{Miz01, Cuo06, Kha13, Akb14, Kun15, Jai17, Kun17, Das18,
Arx23, Zha20, Fel20, Str21}: 
On the one hand, a sufficiently large SOC can rearrange the low-energy levels
of $t_{2g}^4$ ions to form a non-magnetic singlet ground state, separated in
energy from triplet ionic excitations by the SOC value. On the other hand,
superexchange interactions favor a dynamical mixing in this singlet-triplet
level structure, promoting the condensation of triplet excitations to form
long-range antiferromagnetic order. Consequently, if there is right balance
between the singlet-triplet splitting and the strength of the superexchange
interactions, then these two effects are competing and a soft-spin magnetism
emerges~\cite{Kha13,Jai17}. 
This results in an unusual excitation spectrum, which features both
magnon-like excitations and pronounced spin-length fluctuations in a form of
amplitude oscillations of the condensate (an amplitude or Higgs
mode)~\cite{Jai17,Sou17}. 
Interestingly, exactly this regime seems to be realized in $4d^4$
\mbox{Ca$_2$RuO$_4$}, being facilitated by a substantial lowering of the
triplet excitations in the presence of a tetragonal crystal field
\cite{Jai17}.

In this paper, we focus on the interplay of the above unusual soft-spin
magnetic background with doped electron-like carriers. In general, doping
Mott insulators remains one of the major challenges in the study of strongly
correlated electron systems~\cite{Lee06, Ima98, Kho10}. Perhaps the only case
that is relatively well-understood is the case of a single hole (or electron)
doped into a single-band Mott insulator. In this case, a doped hole disrupts
the magnetic structure, leading to strong coupling between the hole and
magnetic excitations~\cite{Lee06}. Since the 2D Mott insulator supports
long-range antiferromagnetic (AF) order the hole couples strongly to the
collective excitations of the AF order (magnons), forming a well-defined
magnetic polaron quasiparticle~\cite{Bul68, Sch88, Kan89, Mar91, Man07, Gru18,
Bie19, Wrz21}. The single-band magnetic polaron concept has been rigorously
tested and observed in the cold-atom simulations and in the ARPES spectra of
copper oxides~\cite{Chi18, Koe19, Koe21, Wel95, Ron05, Bac25}. In contrast,
the situation in Mott insulators with more than one orbital active in the
valence band, both with and without SOC, seems more intricate: On the level of
effective lattice models, earlier works on $3d$ transition metal compounds,
such as manganites~\cite{Kil99}, cobaltates~\cite{Cha07} or
vanadates~\cite{Woh09}, suggested onset of magnetic polarons---of spin or
spin-orbital nature. Similarly, also in systems with non-negligible SOC, an
onset of magnetic polarons was theoretically postulated~\cite{Pae17}. Going
to the more extreme case of Kitaev systems, the dynamic interplay between the
doped carriers and the exotic excitations of the spin-liquid background lead
to the suggestion of electron fractionalization~\cite{Tro13, Hal14, Wan18}.
However, DFT+U calculations often align well with ARPES results for Mott
insulators with orbital degrees of freedom. This happens for the van der Waals
magnet \mbox{MnPS$_3$}~\cite{Str23}, the spin-orbit coupled 2-1-4
iridates~\cite{Kim08, Tor15, Zha13a, Len19} and---even more surprisingly---for
the Kitaev candidate materials~\cite{Com12, Zho16}. Thus it seems as
mean-field approaches are often sufficient to describe the photoemission
spectra of these systems. 

\mbox{Ca$_2$RuO$_4$} clearly illustrates the difficulties mentioned above in
identifying a spin polaron in an AF Mott insulator with active orbital degrees
of freedom and non-negligible SOC. In fact, there is a notable inconsistency
in the approaches used to describe the magnetic versus the electronic
properties of \mbox{Ca$_2$RuO$_4$}: On one side, the magnetic phenomena are
best described by a strongly correlated, localized electron picture. This is
particularly evident in the observation of an amplitude (Higgs) mode in its
inelastic neutron scattering data~\cite{Jai17}, which was successfully
interpreted by employing a complex spin-orbital superexchange
model~\cite{Akb14,Jai17}. In contrast, explanations for the key electronic
properties of \mbox{Ca$_2$RuO$_4$} have largely relied on mean-field
approaches. For example, the measured ARPES spectrum has been explained using
a combination of DFT and single-site DMFT~\cite{Sut17}, where local
correlation effects were fully incorporated and proved crucial for
understanding the onset of Hund’s-induced splitting of the $d_{xy}$ band, as
observed in ARPES. This result was recently confirmed by the GW+EDMFT
approach~\cite{Pet21}. One reason for the relatively good agreement of
mean-field-based approaches with the ARPES data is that the spectra were
measured at temperatures above $T_\mathrm{N}$, which largely suppresses {\it
intersite} correlation effects. Although some weak signatures of spin polaron
physics in the measured ARPES spectra of \mbox{Ca$_2$RuO$_4$} \cite{Sut17}
were discussed in \cite{Klo20}, the model employed there was a largely
phenomenological $S=1$ $t$--$J$ model and solely the response of the
$d_{xy}/d_{yz}$ orbitals was taken into account. 

The aim of the present work is to examine the spin-polaronic features in a
realistic spin-orbital $t$-$J$-like model for \mbox{Ca$_2$RuO$_4$} derived on
microscopic grounds and to understand the implications of the soft-spin nature
of the magnetic background. We choose to study the {\it inverse} photoemission
spectrum, i.e. the propagation of a single extra electron rather than the more
widely discussed single hole / photoemission case. The reason is that
injecting a single electron into a $d^4$ background of Ru$^{4+}$ ions results
in a nominal $d^5$ configuration, which---unlike the $d^3$ configuration
relevant to ARPES---is largely affected by SOC and carries a {\it pseudospin}
$J=\frac12$. 

The paper starts with the derived effective model specified in
Sec.~\ref{sec:model}. Extending largely the earlier works \cite{Cha16,Jai17},
it captures the interactions among low-energy $d^4$ and $d^5$ degrees of
freedom with complex internal structure reflecting the spin-orbital
entanglement due to SOC. In Sec.~\ref{sec:d4backgr} we review the essential
physics of the $d^4$ background, discussing its phases and magnetic excitation
spectra. The central part of the paper consists of Secs.~\ref{sec:SCBA} and
\ref{sec:numerics} presenting a study of the single $d^5$ carrier propagating
through the $d^4$ background. Section~\ref{sec:SCBA} provides the necessary
technical basis in the form of self-consistent Born approximation (SCBA)
equations. These equations are then numerically solved and the results
interpreted in Sec.~\ref{sec:numerics}.
For a better understanding, we examine the behavior of the single carrier as a
function of the relative strength of the tetragonal crystal field as compared
to the SOC. While in \mbox{Ca$_2$RuO$_4$} the ratio of these two parameters is
fixed, changing its value enables us to conveniently drive the system between
its two phases---nonmagnetic (NM) phase built on ionic singlets and
antiferromagnetically (AF) ordered phase associated with a condensate of ionic
triplets. This way the main results are uncovered: In the NM phase, the
carrier {\it seems} to primarily move as a free particle, however, a closer
look reveals that a pseudogap is always present in the quasiparticle
dispersion. This observation is further addressed in Sec.~\ref{sec:toy} with
the help of a simple toy model. In contrast, the carrier motion in the AF
phase is predominantly of polaronic nature, albeit remnants of free electron
motion are still visible in the regime relevant to \mbox{Ca$_2$RuO$_4$}.
A~detailed analysis presented in Sec.~\ref{sec:numerics} reveals that the
observed polaronic motion is far more heavily influenced by the coupling to
the amplitude mode than to the magnon one.


\section{Model}
\label{sec:model}

In this section we describe the microscopic model used to study the
propagation of $d^5$ electron-like carriers in a soft-spin $d^4$ background.
The model is obtained as an effective low-energy limit of a three-orbital
$t_{2g}$-based Hubbard Hamiltonian on a square lattice. We consider the
Hubbard model to be in the Mott regime and including SOC of sufficient strength
so that the latter is decisive for the structure of the effective
low-energy degrees of freedom as will be discussed in Sec.~\ref{sec:basis}. 
The resulting exchange interactions within the $d^4$ background itself and its
interaction with an injected $d^5$ carrier will be exposed in
Secs.~\ref{sec:d4int} and \ref{sec:d5d4int}, respectively.
The values of the model parameters are chosen to describe the canonical
spin-orbit-coupled $d^4$ material \mbox{Ca$_2$RuO$_4$}---though we keep the
tetragonal crystal field as a free control parameter allowing us to examine
how the physics changes once the system is driven through a quantum phase
transition.

\subsection{Low-energy $d^4$ and $d^5$ ionic states}
\label{sec:basis}

\begin{figure}[t!b]
\includegraphics[scale=1.0]{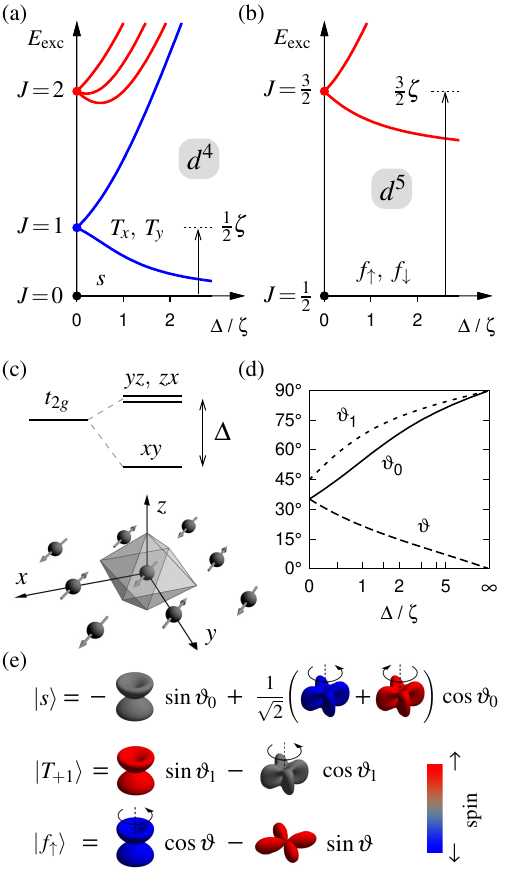}
\caption{{\bf $d^4$ and $d^5$ ionic states:} 
Low-lying energy levels of $t_{2g}^4$ (a) and $t_{2g}^5$ (b) ions as functions
of tetragonal crystal field $\Delta$ measured by the SOC strength $\zeta$. The energies
are plotted relative to the ground-state energy (i.e. $J=0$ or $J=\frac12$
levels, respectively). The vertical scale is common to both plots.
(c)~Splitting among $t_{2g}$ orbitals induced by a tetragonal crystal-field 
$\Delta>0$. Within point-charge model, positive $\Delta$ corresponds to
a compression of metal-O$_6$ octahedra in the direction perpendicular to
the $xy$ plane defined by the square lattice as shown below. 
The particular AF order observed in \mbox{Ca$_2$RuO$_4$} is also indicated.
(d)~Auxiliary angles entering the factors in the ionic wave functions of
Eqs.~\eqref{eq:swf}--\eqref{eq:fwf} and panel (e).
(e)~Schematic picture of selected relevant ionic states: $t_{2g}^4$ ionic ground
state $|s\rangle$, $|T_{+1}\rangle$ representing one of the triplet
excitations, and $|f_{\uparrow}\rangle$ from the $t_{2g}^5$ ground-state pair.
We utilize hole representation taking $t_{2g}^6$ state as a reference. The
components of the two- or one-hole wave functions are depicted as polar plots
of the density distribution with the spin polarization indicated by color. 
Orbital angular momentum is represented by the circular arrows.
}\label{fig:multiplet}
\end{figure}

We start by specifying the basis of our effective model. To this end, we
determine the multiplet structure of $d^4$ and $d^5$ transition metal ions
adopting the $LS$ coupling scheme when incorporating SOC and select the relevant low-energy states. 
Focusing on $4d$ or $5d$ transition metals, all the valence electrons can be
assumed to reside in $t_{2g}$ orbitals and form either spin $S=1$ in case of
$t_{2g}^4$ configuration or spin $S=\frac12$ in case of $t_{2g}^5$
configuration.
Let us first consider the case of crystal field of cubic
symmetry implying degenerate $t_{2g}$ orbitals. 
In such situation, both $t_{2g}^4$ and $t_{2g}^5$ configurations carry
effective orbital angular momentum $L=1$, which is combined with the spin
by virtue of spin-orbit coupling into the total angular momentum 
$\vc J=\vc L+\vc S$ eigenstates having $J=0,1,2$ 
(i.e. singlet-triplet-quintuplet level structure) or $J=\frac12,\frac32$ 
(doublet-quartet), respectively~\cite{Tak21}.
We now add a tetragonal crystal-field component $\Delta$ that is generated
e.g. by an out-of-plane contraction/elongation of the metal-O$_6$ octahedra. 
When including this crystal field, the $t_{2g}$
orbitals split as indicated in Fig.~\ref{fig:multiplet}(c) and consequently
the above states get modified with their degeneracy being partially lifted.
The resulting energy levels are presented in Figs.~\ref{fig:multiplet}(a),(b)
as functions of $\Delta$ measured by the single-electron spin-orbit coupling
$\zeta$. We consider only positive $\Delta$ as realized in
\mbox{Ca$_2$RuO$_4$} (in this case $\Delta/\zeta$ was estimated to $\approx 1.5$ in
Ref.~\cite{Jai17}).

Using the effective model, we aim to address the low-energy window and
the regime $\Delta\gtrsim\zeta$, we thus omit:
\textit{(i)}~the $d^4$ states derived from $J=2$, 
\textit{(ii)}~one of the original $J=1$ triplets that 
is pushed up in energy by positive $\Delta$, and 
\textit{(iii)}~the high-energy $d^5$ states. The remaining states included
in our basis can be compactly expressed in terms of $|L^z S^z\rangle$ states
as follows: 

{\it First}, we consider three $d^4$ eigenstates. The singlet $d^4$ ground state takes the form
\begin{equation}\label{eq:swf}
|s\rangle = 
- \sin\vartheta_0\, |\,0,0\rangle
+\cos\vartheta_0\, \tfrac1{\sqrt2}\bigl(|+\!1,-\!1\rangle+|-\!1,+\!1\rangle\bigr) 
\end{equation}
with the parameter $\vartheta_0$ determined by $\delta=\Delta/\zeta$ via 
$\tan2\vartheta_0 = 2\sqrt2/(1-2\delta)$. The relevant two members of the
original $J=1$ $d^4$ triplet that got lowered in energy by positive $\Delta$
are based on $J^z=\pm 1$ eigenstates
\begin{equation}\label{eq:Twf}
|T_{\pm 1}\rangle = 
\pm\sin\vartheta_1\, |\,0,\pm1\rangle
\mp\cos\vartheta_1\, |\pm1,0\rangle 
\end{equation}
with $\vartheta_1$ satisfying $\tan\vartheta_1 =
1/(\sqrt{1+\delta^2}-\delta)$. For later convenience, we will utilize their linear combinations
$|T_x\rangle=\frac{i}{\sqrt2}(|T_{+1}\rangle-|T_{-1}\rangle)$
and
$|T_y\rangle=\frac{1}{\sqrt2}(|T_{+1}\rangle+|T_{-1}\rangle)$.

{\it Second}, we consider the $d^5$ doublet
\begin{align}
|f_\uparrow\rangle &= \cos\vartheta\, |+\!1,\downarrow\rangle 
	             -\sin\vartheta\, |\,0,\uparrow\rangle \,, \notag\\
|f_\downarrow\rangle &= \sin\vartheta\, |\,0,\downarrow\rangle 
	               -\cos\vartheta\, |-\!1,\uparrow\rangle \label{eq:fwf}
\end{align}
carrying \mbox{pseudospin-$\frac12$}.
The parameter $\vartheta$ is determined by $\tan 2\vartheta = 2\sqrt{2}/(1+2\delta)$.

We also give an explicit expression for the energy splitting between
$|s\rangle$ and $|T_{x,y}\rangle$ states, denoted here as
$E_T=E(T_{x,y})-E(s)$, as the most crucial parameter of our model
arising from ionic physics:
\begin{equation}\label{eq:SMETps1}
E_T= \tfrac14\zeta\left[1+\sqrt{(1-2\delta)^2+8}-2\sqrt{1+\delta^2}\right] \;.
\end{equation}
Note that the $E_T$ splitting is largely reduced from its $\Delta/\zeta=0$
value $E_T=\zeta/2$ already at moderate $\Delta\approx\zeta$, which enables 
the exchange interactions to overcome the splitting and induce
the condensation of $T$ as discussed later. 
The splitting vanishes completely at $\Delta/\zeta\rightarrow\infty$.

Pictorial representation of some of the above states shown in
Fig.~\ref{fig:multiplet}(e) emphasizes their complex spin-orbital entangled
nature (as imposed by SOC), which gets reflected in the anisotropy of the exchange
interactions. For example, \mbox{pseudospin-$\frac12$} carried by
$|f_\uparrow\rangle$, $|f_\downarrow\rangle$ doublet is formally captured by
\mbox{spin-$\frac12$}, but due to their internal structure, these pseudospins may be
subject to highly anisotropic interactions such as in the honeycomb Kitaev
systems~\cite{Jac09}. When increasing $\Delta/\zeta$ that enters via the three
variables $\vartheta_0$, $\vartheta_1$, $\vartheta$ [see
Fig.~\ref{fig:multiplet}(d)], some components of the wave functions get
gradually suppressed and the complexity stemming from the spin-orbital
entanglement fades away. For instance, in the case of the three $d^4$ states,
the orbital angular momentum gets fully quenched in the
$\Delta/\zeta\rightarrow\infty$ limit and we are left with pure \mbox{spin-1}
situation.


\subsection{Magnetic interactions in the $d^4$ background}
\label{sec:d4int}

Having selected our local (ionic) basis, we continue by specifying the bond
interactions constituting the effective model.  We address first the exchange
interactions within $d^4$ background, that we have derived based on the
square-lattice $t_{2g}$ Hubbard model with SOC.  The resulting $d^4$ part of
the effective model is identical to the model used earlier in
Ref.~\cite{Jai17} to fit the neutron data. Our derivation complements this
model with explicit expressions for the exchange interactions in terms of the
parameters of the underlying Hubbard model with SOC.

The effective $d^4$ model can be conveniently formulated as a hardcore boson
model with the hardcore bosons being assigned to the three local-basis states
$s$, $T_x$ and $T_y$.  Particularly compact exchange expressions are obtained
when introducing \mbox{pseudospin-1} operating in the three-dimensional
Hilbert space spanned by the selected $d^4$ states. Its components expressed
using the hardcore boson operators read as
\begin{align}
\widetilde{S}^x &= -i\,(s^\dagger T^\pd_x - T^\dagger_x s) \,, \label{eq:pseudo1x} \\
\widetilde{S}^y &= -i\,(s^\dagger T^\pd_y - T^\dagger_y s) \,, \label{eq:pseudo1y} \\
\widetilde{S}^z &= -i\,(T^\dagger_x T^\pd_y - T^\dagger_y T^\pd_x) \,. \label{eq:pseudo1z}
\end{align}
The auxiliary \mbox{spin-1} has also a physical meaning---the in-plane components 
$\widetilde{S}^x$, $\widetilde{S}^y$ carry van Vleck magnetic moment while 
the out-of-plane component $\widetilde{S}^z$ corresponds to the magnetic 
moment hosted by the $T$ states. Details on this correspondence and the related 
$g$-factor values can be found in the Supplemental Material of Ref.~\cite{Cha24}.

The exchange model for the Mott limit is derived in the usual way by
perturbatively (to second order) eliminating the nearest-neighbor hopping in
the relevant $t_{2g}$ Hubbard model on the square lattice.  In this case,
there are always two of the three $t_{2g}$ orbitals active for a given bond
direction, for example the hopping along the $x$-bond $\langle ij\rangle$
(i.e. bond parallel to the $x$ axis) takes the form 
\begin{equation}\label{eq:thop}
-t\sum_{s=\uparrow,\downarrow} (d_{xy,s,i}^\dagger d_{xy,s,j}^\pd 
+ d_{zx,s,i}^\dagger d_{zx,s,j}^\pd) + \mathrm{H.c.}
\end{equation}
Similarly, for the $y$-bonds, the orbitals $xy$ and $yz$ are active.
Together with the local terms explained below, the resulting Hamiltonian
that describes the $d^4$ background can be written as
\begin{multline}\label{eq:Hd4}
\mathcal{H}_{d^4} = \sum_{i\in\text{sites}} \left\{ E_T (\widetilde{S}^z_i)^2 
-\tfrac12\Delta'\left[ (\widetilde{S}^\parallel_i)^2
-(\widetilde{S}^\perp_i)^2 \right] \right\} +
\\
+\sum_{\langle ij \rangle\in\text{bonds}} \left[
 J_x\, \widetilde{S}^x_i \widetilde{S}^x_j 
+J_y\, \widetilde{S}^y_i \widetilde{S}^y_j
+J_z\, \widetilde{S}^z_i \widetilde{S}^z_j \right] \,.
\end{multline}

The first local term of Eq.~\eqref{eq:Hd4} captures the splitting of the $s$
and $T_{x,y}$ $d^4$ ionic states discussed in the previous section [see
Eq.~\eqref{eq:SMETps1}].  It can also be cast to the form $E_T n_T$ with
$n_{T}=T^\dagger_{x} T^\pd_{x}+T^\dagger_{y} T^\pd_{y}$ and simply counts the
number of $T$ particles at a given site, penalizing them with the splitting
energy $E_T$. The second local term  of Eq.~\eqref{eq:Hd4} is included {\it
ad-hoc} to account for the in-plane anisotropy observed experimentally in
\mbox{Ca$_2$RuO$_4$}.  It fixes the ordered moment direction in the $xy$ plane
and also opens the magnon gap needed to properly fit the neutron data. The
pseudospin component $\widetilde{S}^\parallel$ points along the ordered moment
direction, in the case of \mbox{Ca$_2$RuO$_4$} along $(x+y)/\sqrt2$ [see
Fig.~\ref{fig:multiplet}(c)], i.e.
$\widetilde{S}^\parallel_i = (\widetilde{S}^x_i+\widetilde{S}^y_i)/\sqrt2$.
The component $\widetilde{S}^\perp$ is the perpendicular one:
$\widetilde{S}^\perp_i = (-\widetilde{S}^x_i+\widetilde{S}^y_i)/\sqrt2$.

\begin{figure}[t!b]
\includegraphics[scale=1.0]{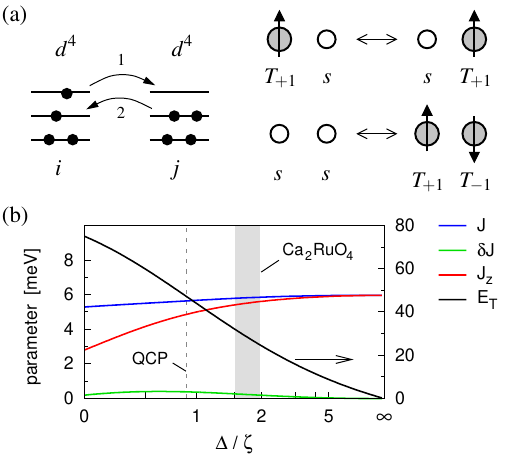}
\caption{ 
{\bf Magnetic interactions in the $d^4$ background:}
(a)~Cartoon representation of important exchange processes in the $d^4$
background. Within second-order perturbation theory, we consider pairs of
virtual electron hoppings that preserve the $d^4$ valence at neighboring sites and
collect them in $\mathcal{H}_{d^4}$ of Eq.~\eqref{eq:Hd4}. The latter connects various
combinations of bond states made of ionic ground-state singlets $s$ and local
$T$ excitations. The essential contributions are the effective hopping of $T$
(top right) and a creation (and similarly an annihilation) of pairs of $T$
(bottom right).
(b)~Parameters entering the $d^4$ magnetic model \eqref{eq:Hd4} as functions
of $\Delta/\zeta$ for fixed $U=3\:\mathrm{eV}$,
$J_\mathrm{H}=0.5\:\mathrm{eV}$, $\zeta=0.15\:\mathrm{eV}$, and
$t=0.143\:\mathrm{eV}$. The scale for $E_T$ is on the right.  Dashed line
marks the phase transition (quantum critical point -- QCP) between the NM and
AFM phases obtained at the level of Eq.~\eqref{eq:d4trial}.  The estimated
$\Delta/\zeta$ ratio for \mbox{Ca$_2$RuO$_4$} is indicated by shading.
}\label{fig:modeld4}
\end{figure}

The exchange parameters $J_x$ and $J_y$ of Eq.~\eqref{eq:Hd4} are
bond-selective and take the values $J_x=J\pm\delta J$, $J_y=J\mp\delta J$ with
the signs depending on the bond direction---upper sign for an $x$-bond, bottom
sign for a $y$-bond. This selectivity is a consequence of spin-orbital
entanglement combined with the orbital selection rules for the hopping
processes. The derived rather lengthy expressions for $J$, $\delta J$, and
$J_z$ in terms of the Hubbard model parameters are given in
Appendix~\ref{app:d4}. Let us note, that in Ref.~\cite{Jai17} the same model
is formulated in a rotated coordinate frame, altering its form. Our parameter
$\delta J$ corresponds to $A$ of Ref.~\cite{Jai17}.  Utilizing the definitions
\eqref{eq:pseudo1x}--\eqref{eq:pseudo1z}, the interactions in
Eq.~\eqref{eq:Hd4} can be converted back to the language of hardcore bosons.
The main exchange processes, associated with $J_x$ and $J_y$, are sketched in
Fig.~\ref{fig:modeld4}(a). They include the effective hopping of $T$ particles
and their creation/annihilation in properly matched pairs.  The bilinear
exchange terms in Eq.~\eqref{eq:Hd4} are accompanied by biquadratic ones, but
their magnitude is found to be negligible.

Figure~\ref{fig:modeld4}(b) shows the evolution of $d^4$ model parameters with
varying $\Delta/\zeta$. It was constructed by fixing the ionic
parameters at the values appropriate for Ru ions in \mbox{Ca$_2$RuO$_4$}:
Hubbard repulsion $U=3\:\mathrm{eV}$, Hund's coupling constant
$J_\mathrm{H}=0.5\:\mathrm{eV}$, and SOC strength $\zeta=0.15\:\mathrm{eV}$.
The hopping amplitude was set to $t=0.143\:\mathrm{eV}$.  These values lead to
a good overall agreement with the estimated $J\approx 5.8\:\mathrm{eV}$,
$E_T\approx 25\:\mathrm{meV}$ obtained by fitting the INS data on
\mbox{Ca$_2$RuO$_4$} \cite{Jai17}, considering the relevant $\Delta/\zeta$
interval of about $1.5-2.0$.  However, as mentioned earlier, we will not limit
ourselves solely to the \mbox{Ca$_2$RuO$_4$} case and use $\Delta/\zeta$ in a
wider range as a convenient handle to drive the model through different
regimes.  When applying such a virtual ``straining'' of the crystal, we will
adopt the effective parameter values given in Fig.~\ref{fig:modeld4} (and
later include those of Fig.~\ref{fig:modeld5d4}).  As observed in
Fig.~\ref{fig:modeld4}(b), the variations in $\Delta/\zeta$ mainly modify the
value of $E_T$, which gets reduced with increasing $\Delta/\zeta$, enabling
the exchange interactions to induce a condensation of $T$ particles at 
a certain point. In the extreme limit $\Delta/\zeta\rightarrow\infty$, the exchange
reaches the isotropic Heisenberg form and the $s$ and $T_{x,y}$ levels merge.
In this case the orbital angular momentum gets fully quenched and we deal with
a real \mbox{spin-1} model.


\subsection{Propagation of doped $d^5$ electron-like carriers \\ in the $d^4$ background}
\label{sec:d5d4int}

\begin{figure}[t!b]
\includegraphics[scale=1.0]{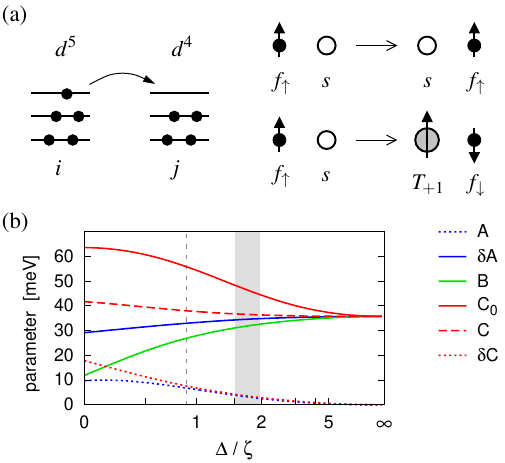}
\caption{
{\bf Propagation of $d^5$ states in the $d^4$ background:}
(a)~Motion of doped electron-like $d^5$ carriers in the $d^4$ background
originates in the hopping processes between the $d^5$ and $d^4$ ions.
These may lead, among other possibilities, to a simple motion without
any pseudospin change (top right) or generate/absorb an excitation in
the background accompanied by a simultaneous change of pseudospin
state of the doped carrier (bottom right).
(b)~Parameters entering the $d^5$--$d^4$ part of the model \eqref{eq:Hd5d4} 
calculated for the same setup as in Fig.~\ref{fig:modeld4}(b).
}\label{fig:modeld5d4}
\end{figure}

By introducing extra electrons into the $d^4$ system, we get a mixture of
$d^4$ and $d^5$ ions with the $d^5$ configuration acting as a mobile
electron-like carrier. The latter essentially moves via the regular
nearest-neighbor hopping $t$ as given by Eq.~\eqref{eq:thop}. When deriving the
low-energy model, this motion has to be expressed in the low-energy basis
composed of $s$, $T_{x,y}$, and $f_{\uparrow,\downarrow}$ states introduced in
Sec.~\ref{sec:basis}.  The resulting $d^5$--$d^4$ coupling Hamiltonian
encompasses the motion of doped $d^5$ carriers represented by fermionic $f$
particles and a simultaneous ``counterflow'' in the $d^4$ background involving
the $s$ and $T_{x,y}$ bosons.
Due to the non-trivial structure of the background, we get a number a
contributions to the $f$ propagation---apart from the usual hopping
insensitive to the \mbox{pseudospin-$\frac12$} carried by $f$, there are also
pseudospin-selective hoppings as well as pseudospin-flip terms [see
Fig.~\ref{fig:modeld5d4}(a) for an example], all of these accompanied by a
properly arranged $d^4$ counterflow.  A detailed description of the individual
processes and their role will be given in Secs.~\ref{sec:scatanal} and
\ref{sec:intuitive} when analyzing the actual numerical results on $f$
propagation.

To efficiently write down the $d^5$--$d^4$ coupling, we introduce a bond
analog of the \mbox{pseudospin-$\frac12$} carried by the doped electron, 
defined as
\begin{equation}\label{eq:bondsigma}
\sigma_{ij}^{\nu} = \sum_{s,s'=\uparrow,\downarrow} 
f^\dagger_{si} \,\sigma_{ss'}^{\nu}\, f^{\phantom{\dagger}}_{s'j}
\,,
\end{equation}
where $\sigma^\nu$ of the right hand side stands for the Pauli matrices. The
$\nu=0$ component of the bond operator $\sigma_{ij}$ containing the identity
matrix $\sigma^0$ corresponds to the regular hopping of the form
$\sigma_{ij}^0 = 
(f^\dagger_{\uparrow i} f^{\phantom{\dagger}}_{\uparrow j} +
f^\dagger_{\downarrow i} f^{\phantom{\dagger}}_{\downarrow j})$, the others
to the pseudospin-selective hopping ($\sigma_{ij}^z$) and to pseudospin flips
($\sigma_{ij}^{x,y}$).
Similarly, the counterflow in the $d^4$ background is best expressed using
bond analogs of the \mbox{pseudospin-1} operators:
\begin{equation}\label{eq:bondSxy}
\widetilde{S}_{ji}^\nu = 
-i\,(s^\dagger_j T^\pd_{\nu i} - T^\dagger_{\nu j}s^\pd_i)
\,,\qquad(\nu=x,y)
\end{equation}
and
\begin{equation}\label{eq:bondSz}
\widetilde{S}_{ji}^z =
-i\,( T^\dagger_{x j} T^\pd_{y i} - T^\dagger_{y j} T^\pd_{x i} ) \,,
\end{equation}
which have the same structure as in
Eqs.~\eqref{eq:pseudo1x}-\eqref{eq:pseudo1z} but the two bosonic operators are
now assigned to the two sites of a nearest-neighbor bond.  Altogether, the
$d^5$--$d^4$ coupling, which splits naturally into a spin-spin and
density-density channel, reads as
\begin{multline}\label{eq:Hd5d4}
\mathcal{H}_{d^5\text{--}d^4} = 
- \sum_{\langle ij\rangle} \left\{
A_x\, \sigma_{ij}^{x} \widetilde{S}_{ji}^x +
A_y\, \sigma_{ij}^{y} \widetilde{S}_{ji}^y -
B\, \sigma_{ij}^{z} \widetilde{S}_{ji}^z \right. + \\
\left.
+\sigma^0_{ij}
\left[ C_0\, s^\dagger_j s^\pd_i + 
C_x\, T^\dagger_{x j} T^\pd_{x i} + 
C_y\, T^\dagger_{y j} T^\pd_{y i} \right]
+\text{H.c.}\right\} \,.
\end{multline}
As in the case of Eq.~\eqref{eq:Hd4}, the $x$ and $y$ components 
of interaction parameters are bond-selective, namely
$A_x=A\pm\delta A$, $A_y=A\mp\delta A$ and
$C_x=C\mp\delta C$, $C_y=C\pm\delta C$
with the upper and lower signs being applied in the case of $x$ and $y$ bonds,
respectively. The signs in Eq.~\eqref{eq:Hd5d4} are adjusted in such a way
that all the parameters are positive in the whole $\Delta/\zeta>0$ range.

The values of the interactions are shown in Fig.~\ref{fig:modeld5d4}(b),
assuming again the microscopic parameter setup appropriate for
\mbox{Ca$_2$RuO$_4$} as in Fig.~\ref{fig:modeld4}(b).  The corresponding
expressions are summarized in Appendix~\ref{app:d5d4} (the interactions in a
simplified form for the $\Delta/\zeta=0$ case can be also found in the initial
study \cite{Cha16}).  Interestingly, despite the small bond-selectivity found
in the case of $d^4$ Hamiltonian, here it is found to dominate the
pseudospin-flip $A$ terms.  In the $\Delta/\zeta\rightarrow\infty$ limit, the
states $s$, $T_{x,y}$ are degenerate and participate equally in the
$d^5$--$d^4$ processes, leading to $|A_x|=|A_y|=B=C_0=C$. Finally, let us
comment on the trivial case of the nonmagnetic background composed
exclusively of $s$ singlets with $T$ particles completely suppressed.  In such
a case, obtained in large $E_T$ limit, the $d^5$--$d^4$ coupling reduces to
only a ``simple'' hopping contained in the $\propto C_0$ term with $C_0$ being
the hopping amplitude.



\section{Magnetism of $d^4$ background:\\ phases and excitations}
\label{sec:d4backgr}

In this paragraph we review the phases and unusual excitations hosted by the
(soft-spin) $d^4$ background driven by Hamiltonian~\eqref{eq:Hd4}. The $d^4$
model will be solved on the level of linear flavor-wave theory employed also
in earlier studies \cite{Akb14,Jai17}.  In general this approach is
reminiscent of the linear spin-wave theory, for it expresses the local states
in terms of bosons and handles the bond interactions among them and
constraints in a similar approximative way.  It was initially developed for
anisotropic \mbox{spin-1} systems \cite{Pap84}, later applied also to
\mbox{spin-1} nematics \cite{Pap88,Chu90} as well as in other situations, e.g.
to bilayer Heisenberg magnets \cite{Som01} that can be captured by
singlet-triplet models within the bond operator approach~\cite{Sac90,Chu95}.
In Ref.~\cite{Zha13b}, the method has been used to study phases and
excitations of a 2D $S=1$ spin system with planar anisotropy, formally similar
to our $d^4$ model \eqref{eq:Hd4}, and successfully benchmarked against
quantum Monte Carlo simulations.  The approximation quite well describes the
transition between the quantum paramagnet at large planar anisotropy and the
ordered planar antiferromagnet for a smaller planar anisotropy.  Moreover, it
captures in a unified way both the excitonic excitations in the quantum
paramagnet as well as the Goldstone magnons and amplitude modes of the planar
antiferromagnet~\cite{Jos15}. Note, however, that this approach has certain
limitations at the \mbox{spin-1} Heisenberg point with vanishing planar
anisotropy~\cite{Jos15} (very large $\Delta/\zeta$ limit in our case).

\begin{figure}[t!b]
\includegraphics[scale=1.0]{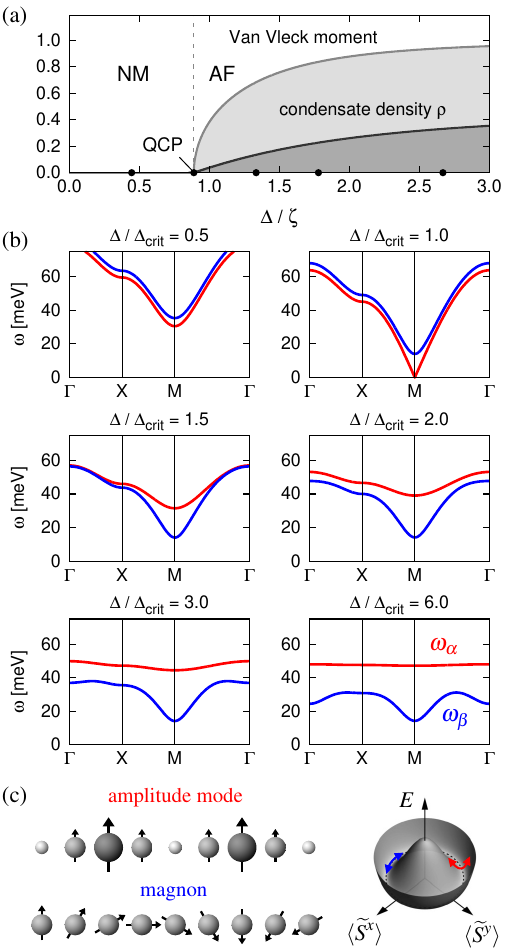}
\caption{
{\bf Ground and excited states of the $d^4$ background:}
(a)~Phase diagram of the $d^4$ model obtained by a variational calculation at
the level of Eq.~\eqref{eq:d4trial} for the parameter values from
Fig.~\ref{fig:modeld4}(b). Above the critical value $\Delta_\mathrm{crit}
\approx 0.9 \zeta$, the initially nonmagnetic system develops an AF order
characterized by van Vleck magnetic moment $\langle \widetilde{S}_\parallel
\rangle$ and $T$-condensate density $\rho$.
(b)~Excitation spectra obtained via linear flavor-wave theory for several
values of $\Delta$ measured relative to its critical value
$\Delta_\mathrm{crit}$.
(c)~Schematic view of the two excitations $\alpha$ and $\beta$.  Upon
excitation by an amplitude mode $\alpha$, the moment length of a given site
oscillates in time, keeping its direction intact. In contrast, the magnon mode
$\beta$ corresponds to pure rotations of the moments.  In the long-wavelength
limit, the modes can also be understood as specific oscillations in the energy
landscape shown on the right (the anisotropy term $\Delta'$ is not considered
in this sketch).
}\label{fig:d4excit}
\end{figure}

Applying the above method to our model of Eq.~\eqref{eq:Hd4}, we first
approximate its ground state by the variationally optimized product state
\begin{equation}\label{eq:d4trial}
|\Psi\rangle = \prod_{\vc R\in\text{sites}} 
\left[
\sqrt{1-\rho}\; s^\dagger + 
\sqrt{\rho}\;\left( 
d^*_x T_x^\dagger + d^*_y T_y^\dagger \right) \right]_{\vc R} |\text{vac}\rangle
\,.
\end{equation}
This trial wave function allows for a condensate of $T$ particles, having a
density $\rho$ and the internal structure captured by the site-dependent
complex vectors $\vc d=(d_x,d_y)$. The simple form of \eqref{eq:d4trial}
transparently highlights the essential feature of our model---the possibility
of a coherent on-site mixing of $s$ and $T$.  By minimizing the average of
$\mathcal{H}_{d^4}$ in the variational ground state \eqref{eq:d4trial}, taking
$\rho$ and the set of $\vc d$ as variational parameters, the model is found to
support two phases. At large $E_T$ compared to the exchange interactions, a
nonmagnetic phase with $\rho=0$ is realized, i.e. the $d^4$ background state
is made of ionic ground states $s$. Once the exchange interactions reach the
critical strength $J_\mathrm{crit} = \frac18(E_T-\frac12\Delta')$, the system
switches---via the condensation of $T$ bosons---to an antiferromagnetic phase with
ordered in-plane van Vleck moments $\langle \widetilde{S}^{x,y}\rangle$,
hosted by on-site superposition of $s$ and $T_{x,y}$ [see
Eqs.~\eqref{eq:pseudo1x} and \eqref{eq:pseudo1y}].  In the latter case 
\begin{equation}
\vc d^*=i\eeQR(\cos\phi,\sin\phi) \,, 
\end{equation}
where $\vc Q$ is the AF ordering vector $\vc Q=(\pi,\pi)$ and $\phi$ is the
angle of the ordered moments in the $xy$-plane that is fixed by $\Delta'$
term. Taking \mbox{Ca$_2$RuO$_4$} as an example, in the following we choose
the corresponding $\phi=\pi/4$. The condensate density obtained by
minimization takes the value $\rho=\frac12[1-(E_T-\frac12\Delta')/8J]$ and
grows with increasing exchange strength (or decreasing $E_T$) up to
$\rho=\frac12$ in the $J/E_T\rightarrow \infty$ limit.
The phase diagram constructed by employing the model parameters as given in
Fig.~\ref{fig:modeld4}(b) is presented in Fig.~\ref{fig:d4excit}(a).  For
simplicity, we ignore the minor parameter $\delta J$. The system is tuned by
varying $\Delta/\zeta$, which influences mainly $E_T$ that becomes
significantly reduced and reaches the critical value at about
$\Delta/\zeta\approx 0.9$. 

To determine the excitation spectra, $\mathcal{H}_{d^4}$ is first transformed
by rotating the bosonic operators $(s,T_x,T_y)$ to a new set $(a,b,c)$ 
following the implicit relations
\begin{align}
s_{\vc R} &= 
   \cos\theta\; c_{\vc R} + i\eeQR\sin\theta\, a_{\vc R} \,, \notag  \\
T_{a\vc R}  &= 
   i\eeQR\sin\theta\, c_{\vc R} + \cos\theta\, a_{\vc R} \,, \notag \\
T_{b\vc R} &= 
   b_{\vc R} \label{eq:bosrot}
\end{align}
with the linear combinations
$T_a=\cos\phi\; T_{x} +\sin\phi\; T_{y}$
and
$T_b=-\sin\phi\; T_{x} +\cos\phi\; T_{y}$
adjusted to match the in-plane \mbox{pseudospin-1} components 
$\widetilde{S}^\parallel$ (parallel to the ordered moment direction) 
and $\widetilde{S}^\perp$ (perpendicular to it) 
appearing in Eq.~\eqref{eq:Hd4}.
The main rotation angle $\theta$ is determined by the condensate density via
$\sin\theta = \sqrt\rho$.
The rotation is designed so that the new bosons play physically distinct
roles. Boson $c$ is associated with the condensate as seen by obtaining 
$|\Psi\rangle=\prod_{\vc R} c_{\vc R}^\dagger\; |\mathrm{vac}\rangle$ 
after the rotation; on the other hand, bosons $a$ and $b$ are associated with
two different kinds of elementary excitations in the system.
The linear flavor-wave expansion is then performed by substituting
$c,c^\dagger \rightarrow \sqrt{1-n_a-n_b}\approx 1-\frac12(n_a+n_b)$ 
to account for the hardcore constraint in a dynamic manner, and collecting
terms up to second order in $a$ and $b$. The result is a sum of two quadratic
Hamiltonians, each of them separately involving either $a$ or $b$ bosons. At
this point neglecting the $\delta J$ term is very convenient, because such a
term would lead to a mixing of $a$ and $b$ modes. Both $a$ and $b$
contributions are diagonalized by successive Fourier and Bogoliubov transformations,
giving rise to $\alpha$ and $\beta$ eigenmodes described by the Hamiltonian
$\sum_{\vc q} \left(
\omega_{\alpha \vc q}\, \alpha_{\vc q}^\dagger \alpha_{\vc q}^\pd +
\omega_{\beta \vc q}\, \beta_{\vc q}^\dagger \beta_{\vc q}^\pd \right)$.
At the level of linear flavor-wave theory, they appear as sharp excitations
with infinite lifetime, a damping would occur when going beyond the quadratic 
expansion \cite{Jai17}. The dispersions of the eigenmodes are given by
$\omega_{\alpha \vc q} = \sqrt{A_{a\vc q}^2-B_{a\vc q}^2}$
and
$\omega_{\beta \vc q} = \sqrt{A_{b\vc q}^2-B_{b\vc q}^2}$,
with the subfactors 
\begin{align}
A_{a\vc q} &= E_T-\tfrac12\Delta'+4J\gamma_{\vc q} \,, \label{eq:ABfactPM1} \\
A_{b\vc q} &= E_T+\tfrac12\Delta'+4J\gamma_{\vc q} \,, \\
B_{a\vc q} &= B_{b\vc q} = 4J\gamma_{\vc q}
\end{align}
valid for the NM phase or 
\begin{align}
A_{a\vc q} &= 4J \left( 2 + \gamma_{\vc q} \cos^2 2\theta \right) \,, \\
B_{a\vc q} &= 4J \gamma_{\vc q} \cos^2{2\theta} 
\,, \\
A_{b\vc q} &= 4J(2+\gamma_{\vc q})\cos^2\theta -4J_z\gamma_{\vc q}\sin^2\theta + \Delta'
\,, \\
B_{b\vc q} &= 4(J \cos^2\theta+J_z \sin^2\theta)\gamma_{\vc q} \label{eq:ABfactAF4} 
\end{align}
to be used in the case of the AF phase.
Here $\gamma_{\vc q}$ denotes the nearest-neighbor factor
$\gamma_{\vc q}=\frac12(\cos q_x + \cos q_y)$ for the square lattice.

The evolution of the excitation spectra with increasing value of
$\Delta/\zeta$---and thus at different points of the phase diagram---is
presented in Fig.~\ref{fig:d4excit}(b). Starting deep in the NM phase, the two
excitations are seen as nearly degenerate (the small splitting is due to the
$\Delta'$ anisotropy). When the critical point is approached, the excitations
soften at the AF momentum $\vc Q$ with the $\alpha$ mode touching zero level
at the critical value of $\Delta/\zeta$. The further evolution (past the
critical point) of the two modes is very different.  The $\alpha$ mode hardens
and becomes a flat high-energy excitation at large $\Delta/\zeta$. The $\beta$
mode remains dispersive, gradually softening around $\vc q=(0,0)$, and keeping
a nearly constant gap at $\vc Q$ of about $\sqrt{8J\Delta'}$. This distinct
behavior can be interpreted based on the physical picture of the two modes
provided by Fig.~\ref{fig:d4excit}(c). The mode $\alpha$ related to boson $a$
corresponds to oscillations of the balance between $s$ and $T$, i.e. the
amplitude of the condensate, which explains its increased stiffness as the
condensate develops and becomes more robust at larger $\Delta/\zeta$. The mode
$\beta$ originating from $b$ bosons can be understood as a magnon gapped by
the in-plane anisotropy $\Delta'$.  The softening of its dispersion around
$\vc q=(0,0)$ reflects the progression towards Heisenberg-like situation due
to vanishing single-ion anisotropy $E_T$ contained in the Hamiltonian
$\mathcal{H}_{d^4}$.


\section{A $d^5$ carrier in the $d^4$ background: \\formal matters}
\label{sec:SCBA}

After briefly inspecting the features of the unusual magnetic background, we
focus on the interactions of a doped electron $f$ (corresponding locally to a
$d^5$ ion) with the elementary magnetic excitations of the background. The
resulting polaronic behavior of the doped electron is captured within the
self-consistent Born approximation (SCBA)~\cite{Mar91, Pae17, Klo20}. The
required $d^5$--$d^4$ interaction Hamiltonian is obtained by starting with
$\mathcal{H}_{d^5\text{--}d^4}$ of Eq.~\eqref{eq:Hd5d4} and applying the same
steps as in the linear flavor-wave expansion, i.e. bosonic rotation
\eqref{eq:bosrot} with a replacement of the condensed boson $c$ by
$\sqrt{1-n_a-n_b}$, followed by an expansion in the $a$ and $b$ bosons and
subsequent Fourier transformation. Finally, the magnetic bosons $a$ and $b$
are expressed in terms of the eigenmodes $\alpha$ and $\beta$ as 
$a_{\vc q} = u_{\alpha\vc q}\alpha^\pd_{\vc q} 
+ v_{\alpha\vc q} \alpha^\dagger_{-\vc q}$
with the Bogoliubov factors
\begin{equation}\label{eq:Bogol}
u_{\alpha\vc q}=\frac1{\sqrt2}\sqrt{\frac{A_{a\vc q}}{\omega_{\alpha\vc q}}+1}
\,, \quad
v_{\alpha\vc q}=\frac1{\sqrt2}\sqrt{\frac{A_{a\vc q}}{\omega_{\alpha\vc q}}
-1}\,\;\mathrm{sgn}\,B_{a\vc q}
\end{equation}
and similar for $b$ and $\beta$. The $A$ and $B$ factors are listed in
Eqs.~\eqref{eq:ABfactPM1}--\eqref{eq:ABfactAF4}.  Two parts of the resulting
coupling Hamiltonian are of interest here:
\textit{(i)}~Zeroth-order terms in $\alpha$ and $\beta$ corresponding to a
``simple'' electron hopping without any excitations of the background
involved. However this hopping is still affected by the very presence of the
magnetic condensate blocking the $f$ motion.
\textit{(ii)}~Terms linear in $\alpha$ and $\beta$ providing the basic
coupling of the magnetic excitations to the doped electrons.
In the following, we discuss the latter coupling in detail and describe how it
gets incorporated into the SCBA scheme.  We consider first the simple yet
instructive case of the nonmagnetic phase before moving on to the
antiferromagnetic one, which is of our main interest but leads to a rather
complex formal structure.

\subsection{SCBA equations for the nonmagnetic phase}
\label{sec:SCBAnm}

In the nonmagnetic phase, the elementary excitations $\alpha$, $\beta$ are
associated with the $T$ particles themselves and the relevant zeroth-
and first-order processes are therefore directly those illustrated in 
Fig.~\ref{fig:modeld5d4}(a).

Zeroth-order interaction term, which does not involve any $T$, corresponds to
a ``free'' $f$ hopping through the nonmagnetic $d^4$ background composed of
$s$. According to Eq.~\eqref{eq:Hd5d4}, its amplitude equals to $C_0$, which
leads to the free motion given by Hamiltonian
\begin{equation}\label{eq:Hd5d4zero}
\mathcal{H}^{(0)}_{d^5\text{--}d^4} = \sum_{\vc k s} \varepsilon_{\vc k} 
f^\dagger_{\vc k s} f^\pd_{\vc k s}
\end{equation}
with the bare dispersion relation of a doped electron:
\begin{equation}\label{eq:baredispPM}
\varepsilon_{\vc k}=-2C_0(\cos k_x+\cos k_y) \,.
\end{equation}

First-order terms, on the other hand, involve a creation or annihilation of
the $T$ excitations.  Since $T_0$ has been removed by the tetragonal crystal
field $\Delta$ and the active $T_{x,y}$ are based on $T_{\pm 1}$, the
corresponding processes always include a flip of the electron
\mbox{pseudospin-$\frac12$} as depicted in Fig.~\ref{fig:modeld5d4}(a). If,
for the sake of brevity, we ignore the $\Delta'$ in-plane anisotropy in this
paragraph, the $\alpha$ and $\beta$ excitations are degenerate, with a common
dispersion
$\omega_{\alpha\vc q}=\omega_{\beta\vc q}=\omega_{\vc q}$
and identical Bogoliubov factors $u_{\alpha\vc q}=u_{\beta\vc q}=u_{\vc q}$, 
$v_{\alpha\vc q}=v_{\beta\vc q}=v_{\vc q}$.
With these simplifications, the first-order coupling of doped electrons $f$ to
the magnetic excitations $\alpha$, $\beta$ is given by
\begin{equation}\label{eq:HintPM}
\mathcal{H}^{(1)}_{d^5\text{--}d^4} = 
\sum_{\vc k\vc q s}
(M_{\vc k\vc q s}^\alpha \alpha^\dagger_{\vc q} + 
 M_{\vc k\vc q s}^\beta \beta^\dagger_{\vc q} ) 
\, f^\dagger_{\vc k-\vc q,-s} f^\pd_{\vc k s}
+\mathrm{H.c.}
\end{equation}
with the matrix element
\begin{equation}
M_{\vc k\vc q s}^{\alpha, \beta} \;\propto\;
\delta A\, (\eta_{\vc k-\vc q} u_{\vc q} - \eta_{\vc k} v_{\vc q}) 
\pm A\, (\gamma_{\vc k-\vc q} u_{\vc q} - \gamma_{\vc k} v_{\vc q}) \,.
\end{equation}
Here, the $\pm$ sign applies to $\alpha$ and $\beta$, respectively, and the
omitted proportionality factor equals $-4i$ for $\alpha$, $-4$ for $\beta$ and
$s=\uparrow$, and $+4$ for $\beta$ and $s=\downarrow$. The geometry of the
hopping enters the matrix element via \mbox{$s$-wave} and \mbox{$d$-wave}
nearest-neighbor formfactors for the square lattice:
$\gamma_{\vc k}=\frac12(\cos k_x + \cos k_y)$ and
$\eta_{\vc k}=\frac12(\cos k_x - \cos k_y)$.

\begin{figure}[t!b]
\includegraphics[scale=1.0]{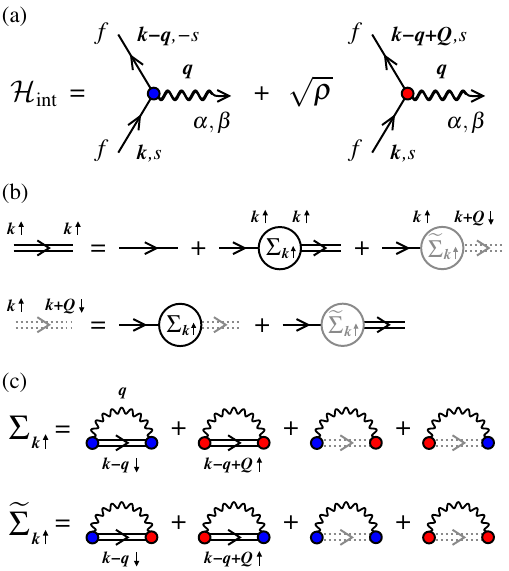}
\caption{
(a)~Diagrammatic representation of the coupling Hamiltonians \eqref{eq:HintPM}
and \eqref{eq:HintAF}. There is always a term flipping pseudospin of the doped
electron upon emission or absorption of magnetic excitations $\alpha$,
$\beta$. In the AF phase, a new contribution appears that is
pseudospin-conserving but brings $\vc Q=(\pi,\pi)$ momentum shift. This
contribution scales with condensate density as $\sqrt{\rho}$.
(b)~Dyson's equations for the normal propagator of the doped electron and the
anomalous one, involving a pseudospin flip and a simultaneous momentum shift
$\vc Q$. Correspondingly, two kinds of selfenergies enter these Dyson's
equations.  The input/output momenta and pseudospins are indicated at the top
of the diagrams.  The simple line corresponds to the only bare propagator
$\mathcal{G}_0=1/(E-\varepsilon_{\vc k})$.  In the nonmagnetic phase, the
anomalous propagator/selfenergy are absent and the Dyson's equations reduce to
the first line with the last term omitted.
(c)~Selfenergy contributions within SCBA for the AF phase. The wiggly line
represents $\alpha$ and $\beta$ excitations (to be summed over) emitted and
absorbed during electron motion.  The internal electron lines corresponding to
normal propagation are labeled by their momenta and pseudospins. In the
nonmagnetic phase, the only non-vanishing term is the first contribution to
the normal selfenergy $\Sigma$.
}\label{fig:SCBA}
\end{figure}

The renormalized electron propagator 
$\mathcal{G}_{\vc k}(E)=[E-\varepsilon_{\vc k}-\Sigma_{\vc k}(E)]^{-1}$ 
is obtained at the level of SCBA, summarized in Fig.~\ref{fig:SCBA} for the
more general case of the AF phase. In the nonmagnetic phase, the SCBA
selfenergy reduces to the first diagram in Fig.~\ref{fig:SCBA}(c) that
translates to a convolution of the renormalized electron propagator and
propagators of the magnetic excitations. 
Note, that even though the interaction in Eq.~\eqref{eq:HintPM} flips the
electron pseudospin, the balance between emitted and absorbed magnetic
excitations during the electron motion restores the initial pseudospin in the
electron propagator, which is then pseudospin-conserving as well as
pseudospin-independent.
For a convenient evaluation, we decompose the SCBA selfenergy into real and
imaginary parts $\Sigma=\Sigma'+i\Sigma''$.  The imaginary part $\Sigma''$ of
the selfenergy is calculated via
\begin{equation}
\Sigma''_{\vc k}(E) = 
-\pi \sum_{\vc q}
|M_{\vc k\vc q}|^2 \, \mathcal{A}_{\vc k-\vc q}(E-\omega_{\vc q}) 
\end{equation}
with the electron spectral function given by
$\mathcal{A}_{\vc k}(E)=-{\pi}^{-1}\mathrm{Im}\,\mathcal{G}_{\vc k}(E)$
and $|M_{\vc k\vc q}|^2$ denoting summed up $\alpha$ and $\beta$
contributions $|M^\alpha_{\vc k\vc q s}|^2+|M^\beta_{\vc k\vc q s}|^2$:
\begin{equation}
32\bigl[
(\delta A)^2\, (\eta_{\vc k-\vc q} u_{\vc q} - \eta_{\vc k} v_{\vc q})^2
+ A^2\, (\gamma_{\vc k-\vc q} u_{\vc q} - \gamma_{\vc k} v_{\vc q})^2 \bigr]
\,.
\end{equation}
The real part $\Sigma'$ of the selfenergy is subsequently obtained by
Kramers-Kronig transformation
\begin{equation}
\Sigma'_{\vc k s}(E) = \frac{1}{\pi}\, \mathcal{P} 
\int\limits_{-\infty}^{+\infty} \frac{\Sigma''_{\vc k s}(\xi)}{E-\xi}\;\mathrm{d}\xi \,.
\end{equation}
and the steps are repeated until a selfconsistent solution of the selfenergy
equation is found.


\subsection{SCBA equations for the antiferromagnetic phase}
\label{sec:SCBAaf}

The case of the antiferromagnetic phase is formally richer, since the now
nontrivial bosonic rotation \eqref{eq:bosrot} with $\theta\neq 0$ and
$\phi=\pi/4$ brings up a number of new terms to the interaction.  The
contributions of zeroth order in $\alpha$, $\beta$ again give rise to
$\mathcal{H}^{(0)}_{d^5\text{--}d^4}$ of Eq.~\eqref{eq:Hd5d4zero} with the
bare electron dispersion $\varepsilon_{\vc k}$ taking this time the form
\begin{align}
\varepsilon_{\vc k}=
&-2\bigl[C_0-(C_0+C)\sin^2\theta\bigr](\cos k_x+\cos k_y) \notag \\
&-2\delta C\cos2\phi (\cos k_x-\cos k_y) \,. \label{eq:baredispAF}
\end{align}
The second term vanishes in our case of interest due to $\phi=\pi/4$, the
first one shows a reduction of the bare bandwidth compared to the nonmagnetic
phase---the bare motion of the doped electron in the staggered magnetic structure
of the AF condensate being increasingly inhibited with growing condensate density
$\rho=\sin^2\theta$. 

The linear coupling to $\alpha$, $\beta$, represented diagrammatically in
Fig.~\ref{fig:SCBA}(a), has the structure
\begin{multline}\label{eq:HintAF}
\mathcal{H}^{(1)}_{d^5\text{--}d^4} =
\sum_{\vc k\vc q s}
\left[  
(M_{\vc k\vc q s}^\alpha \alpha^\dagger_{\vc q} + 
 M_{\vc k\vc q s}^\beta \beta^\dagger_{\vc q} ) 
\, f^\dagger_{\vc k-\vc q,-s} f^\pd_{\vc k s} +
\right. \\
\left.  
(\bar{M}_{\vc k\vc q s}^\alpha \alpha^\dagger_{\vc q} + 
 \bar{M}_{\vc k\vc q s}^\beta \beta^\dagger_{\vc q} )
\, f^\dagger_{\vc k-\vc q+\vc Q,s} f^\pd_{\vc k s}
\right]+\mathrm{H.c.}
\end{multline}
The matrix elements $M_{\vc k\vc q s}^{\alpha,\beta}$ and $\bar{M}_{\vc k\vc q
s}^{\alpha,\beta}$ are given in Appendix~\ref{app:d5d4fab}. Compared to the
nonmagnetic phase, the coupling is extended by pseudospin-conserving terms,
which include a momentum shift by the AF wave vector $\vc Q=(\pi,\pi)$. The
combination of pseudospin-flipping and pseudospin-conserving processes gives
rise to a more complex structure of Dyson's equations including additionally
anomalous propagators of the type 
$\widetilde{\mathcal{G}}_{\vc k\uparrow} \sim
\langle f^\pd_{\vc k+\vc Q\downarrow} f^\dagger_{\vc k\uparrow} \rangle$, 
see Fig.~\ref{fig:SCBA}(b). They involve normal selfenergy
$\Sigma_{\vc{k}s}(E)$ as well as anomalous selfenergy
$\widetilde{\Sigma}_{\vc{k}s}(E)$, which is associated with both a
pseudospin-flip and $\vc Q$ momentum shift. By solving Dyson's equations in
Fig.~\ref{fig:SCBA}(b), the normal electron propagator is obtained in the form
\begin{equation}
\mathcal{G}_{\vc k\uparrow}
=
\left[
E\!-\!\varepsilon_{\vc k}\!-\!\Sigma_{\vc k\uparrow}
-\frac{\widetilde{\Sigma}_{\vc k\uparrow}
\widetilde{\Sigma}_{\vc k+\vc Q\downarrow}}
{E\!-\!\varepsilon_{\vc k+\vc Q}\!-\!\Sigma_{\vc k+\vc Q\downarrow}}
\right]^{-1} \,,
\end{equation}
where the energy arguments were omitted for compactness of the expression. The
anomalous propagator takes the form
\begin{equation}
\widetilde{\mathcal{G}}_{\vc k\uparrow}
=
\left[
\frac
{(E\!-\!\varepsilon_{\vc k}\!-\!\Sigma_{\vc k\uparrow})
(E\!-\!\varepsilon_{\vc k+\vc Q}\!-\!\Sigma_{\vc k+\vc Q\downarrow})}
{\widetilde{\Sigma}_{\vc k\uparrow}}
-\widetilde{\Sigma}_{\vc k+\vc Q\downarrow}
\right]^{-1} 
\,.
\end{equation}
The selfenergies in SCBA approximation include all combinations of the
interaction vertices as depicted in Fig.~\ref{fig:SCBA}(c).
The $\Sigma''$ part of the normal selfenergy reads as
\begin{multline}\label{eq:selfEn2}
\Sigma''_{\vc k s}(E) = 
-\pi {\sum_{\vc q}} \Bigl[
|M^\alpha_{\vc k\vc q s}|^2 \mathcal{A}_{\vc k-\vc q,-s} +
|\bar{M}^\alpha_{\vc k\vc q s}|^2 \mathcal{A}_{\vc k-\vc q+\vc Q,s} \\
+M^\alpha_{\vc k\vc q s} (\bar{M}^\alpha_{\vc k\vc q s})^* \widetilde{\mathcal{A}}_{\vc k-\vc q,-s} + 
\bar{M}^\alpha_{\vc k\vc q s} (M^\alpha_{\vc k\vc q s})^* \widetilde{\mathcal{A}}_{\vc k-\vc q+\vc Q,s} 
\Bigr]_{E-\omega_{\alpha\vc q}} \\
+\text{$\beta$-terms of the same structure}\,,
\end{multline}
with $\mathcal{A}$ and $\widetilde{\mathcal{A}}$ being the spectral functions
corresponding to the normal and anomalous propagators $\mathcal{G}$ and
$\widetilde{\mathcal{G}}$, respectively, all of them having the energy
argument $E-\omega_{\alpha\vc q}$ ($\alpha$-terms) or $E-\omega_{\beta\vc q}$
($\beta$-terms). Similarly, the $\widetilde{\Sigma}''$ part of the anomalous
selfenergy can be expressed as
\begin{multline}\label{eq:selfEa2}
\widetilde{\Sigma}''_{\vc k s}(E) =
\pi \sum_{\vc q} \Bigl[ 
M^\alpha_{\vc k\vc q s} (\bar{M}^\alpha_{\vc k\vc q,-s})^* 
\mathcal{A}_{\vc k-\vc q,-s} + \\
\bar{M}^\alpha_{\vc k\vc q s} (M^\alpha_{\vc k\vc q,-s})^* 
\mathcal{A}_{\vc k-\vc q+\vc Q,s} + 
M^\alpha_{\vc k\vc q s} (M^\alpha_{\vc k\vc q,-s})^* 
\widetilde{\mathcal{A}}_{\vc k-\vc q,-s} \\
+ \bar{M}^\alpha_{\vc k\vc q s} (\bar{M}^\alpha_{\vc k\vc q,-s})^*
\widetilde{\mathcal{A}}_{\vc k-\vc q+\vc Q,s} 
\Bigr]_{E-\omega_{\alpha\vc q}}
+\text{$\beta$-terms}\rule{0mm}{6mm} \,.
\end{multline}
The selfenergy expressions \eqref{eq:selfEn2} and \eqref{eq:selfEa2} can be
further simplified by using the symmetry relations
$\mathcal{A}_{\vc k s}=\mathcal{A}_{\vc k,-s}$
for the normal components and
$\widetilde{\mathcal{A}}_{\vc k s}=\widetilde{\mathcal{A}}^*_{\vc k,-s}$,
$\widetilde{\mathcal{A}}_{\vc k s}=\widetilde{\mathcal{A}}_{\vc k+\vc Q,s}$
for the anomalous ones.  As in the case of the nonmagnetic phase, the system
of Dyson's equations and selfenergy equations is solved by iterations until
selfconsistency is reached.



\section{A $d^5$ carrier in the $d^4$ background: \\numerical results}
\label{sec:numerics}


\begin{figure*}
\includegraphics[scale=1.0]{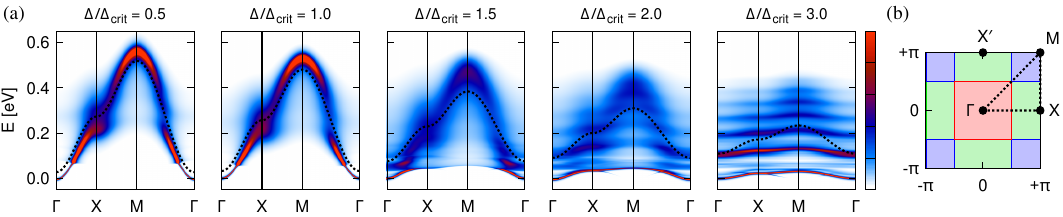}
\caption{{\bf Evolution the electron spectral function across the phase diagram:}
(a)~Spectral function of a single electron propagating in the $d^4$ background
calculated for various values of the control parameter $\Delta$ measured by
its critical value $\Delta_\mathrm{crit}$. The particular parameter points are indicated by black
dots at the $\Delta/\zeta$ scale of the phase diagram presented in
Fig.~\ref{fig:d4excit}(a). The dashed line shows the bare dispersion
$\varepsilon_{\vc k}$
given by Eq.~\eqref{eq:baredispPM} or \eqref{eq:baredispAF}, respectively.
(b)~Brillouin zone of the square lattice with labeled high-symmetry points and
the triangular path used in (a). Color shading defines the partitioning of the
Brillouin zone used in the analysis of the contributing scattering processes
in Fig.~\ref{fig:ImEdecomp}. Keeping in mind the periodic continuation of the
Brillouin zone, there are in total four areas, each associated with
one of the high-symmetry points: $\Gamma$ (red), $M$ (blue), $X$, and $X'$
(both colored as green).
}\label{fig:AkEoverview}
\end{figure*}

Utilizing the SCBA approach described in the previous section, we are able to
study the propagation of doped electrons in the unusual soft-spin background.
The aim of this section is to present the numerical results and analyze the
interplay of doped electrons with the excitations of the magnetic background,
focusing in particular on the AF-ordered phase hosting both amplitude and
magnon excitations. We will contrast the resulting polaronic behavior to spin
polarons thoroughly discussed in context of regular \mbox{spin-$\frac12$}
Heisenberg antiferromagnets.

We start by presenting in Fig.~\ref{fig:AkEoverview}(a) the overall trends in
the normal spectral function $\mathcal{A}_{\vc k}(E)$ through the magnetic
phase diagram.
The parameter values used are those presented earlier in
Figs.~\ref{fig:modeld4}(b) and \ref{fig:modeld5d4}(b), with the exception of
$\delta J$. This bond-selective part of the exchange is omitted to simplify
the excitation spectrum by avoiding the mixing of the amplitude and magnon
excitations, as noted in Sec.~\ref{sec:d4backgr}.
Like in case of the phase diagram shown in Fig.~\ref{fig:d4excit}(a), we have
used the tetragonal crystal field as the main control parameter to drive the
system through the NM/AF phase transition. The locations of the selected
$\Delta/\zeta$ points used in Fig.~\ref{fig:AkEoverview}(a) and covering both
phases are indicated in Fig.~\ref{fig:d4excit}(a) as black dots.

As seen in Fig.~\ref{fig:AkEoverview}(a), the spectra of 
$\mathcal{A}_{\vc k}(E)$ are rather distinct in the NM phase at
$\Delta\leq\Delta_\mathrm{crit}$, where the quasiparticle essentially keeps
tracking the bare dispersion, and in the AF phase at larger $\Delta$, where
clear polaronic features appear. The evolving shape of the spectra with
increasing $\Delta$ has a complex origin and is determined by a combination of
several factors including a reduced bare bandwidth, varying strength of
various scattering processes generated by the several channels contained in
$\mathcal{H}^{(1)}_{d^5\textbf{--}d^4}$ of Eq.~\eqref{eq:HintAF}, and changes
in the magnetic excitation spectrum observed in Fig.~\ref{fig:d4excit}(b).
As we will analyze in detail later, the contrast between NM and AF phases
originates mainly in the rapid development of the condensate after $\Delta$
rises above $\Delta_\mathrm{crit}$. The presence of the condensate
\textit{(i)} boosts the suppression of the bare bandwidth and \textit{(ii)}
opens an additional scattering channel involving the amplitude mode of the
condensate. As a consequence, the ratio (scattering rate)/(bare bandwidth)
steeply increases after entering the AF phase and causes the dramatic changes
observed in $\mathcal{A}_{\vc k}(E)$. A typical ladder-like structure is then
gradually formed when progressing towards larger $\Delta$, this trend is also
supported by the increased flatness of the excitation spectrum, namely the
amplitude mode [see Fig.~\ref{fig:d4excit}(b)]. However, at the same time
$\mathcal{A}_{\vc k}(E)$ keeps showing a dual nature of the electron
motion---both polaronic as well as a free motion, which manifests itself by
remnants of the free dispersion being still well noticeable deep in the AF
phase. In Sec.~\ref{sec:toy}, we will discuss the interplay of these two
components of the electron motion using a simple toy model. 

Before analyzing the details of electron propagation, two general comments are
in order. 
First, even though the calculations were performed using the complete SCBA
scheme as described in Sec.~\ref{sec:SCBAaf}, we found the anomalous
components to have a negligible impact, hence we focus solely on the normal
ones in the following presentation.
Second, let us remind the reader, that the model was constructed primarily for
the low-energy window $\lesssim 0.2\:\mathrm{eV}$ and misses e.g. the minor
contributions from the $d^5$ states based on $J=\frac32$ quartet. The
motivation here is to understand the main effects related to the unusual
excitation spectrum of the magnetic background, in particular the role of the
amplitude mode versus magnons.


\subsection{Relative contributions of the amplitude mode and magnons}

\begin{figure}[t!b]
\includegraphics[scale=1.0]{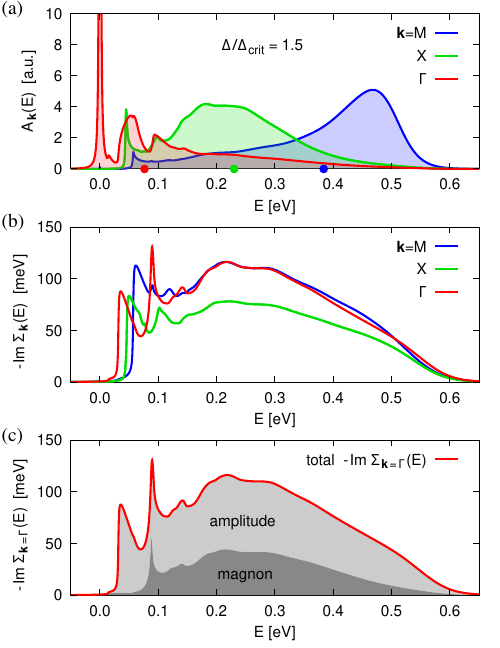}
\caption{{\bf Electron renormalization in the AF phase:}
(a)~Electron spectral function at high-symmetry \mbox{$\vc k$-points} in the
Brillouin zone: $\Gamma=(0,0)$, $X=(\pi,0)$, and $M=(\pi,\pi)$. The parameter
point $\Delta/\Delta_\mathrm{crit}=1.5$ roughly corresponds to the
\mbox{Ca$_2$RuO$_4$} fit given by Ref.~\cite{Jai17}. The colored dots at the
horizontal axis show the bare band energies $\varepsilon_{\vc k}$ for the
individual \mbox{$\vc k$-points}. The spectra were broadened by lorentzians
with FWHM of $1\:\mathrm{meV}$.
(b)~The corresponding imaginary part of the electron selfenergy determining
the scattering rate (inverse lifetime). The step-like increase of
\mbox{$-\mathrm{Im}\,\Sigma$} at low energies correlates with the onset of
damping seen in $\mathcal{A}_{\vc k}(E)$ of panel (a).
(c)~Decomposition of the $\vc k=\Gamma$ selfenergy into the contribution of
the amplitude mode $\alpha$ and that of the magnon $\beta$. The contributions
are represented by the shaded areas that sum up to the total selfenergy.  At
this parameter point, the amplitude-mode contribution has about two times
larger weight than the magnon one and fully dominates at low energies.
}\label{fig:AkEcuts}
\end{figure}

To begin the detailed investigation of the interplay of the magnetic excitations with
the doped carriers, we first focus on a representative point
$\Delta/\Delta_\mathrm{crit}=1.5$ in the AF phase, which roughly corresponds to
the \mbox{Ca$_2$RuO$_4$} case. The relevant $\mathcal{A}_{\vc k}(E)$ map is
included as a middle panel in Fig.~\ref{fig:AkEoverview}(a) and in
Fig.~\ref{fig:AkEcuts}(a) we complement it by \mbox{constant-$\vc k$} cuts of
$\mathcal{A}_{\vc k}(E)$ at high-symmetry points in the Brillouin zone, which
show in detail the quasiparticle features at low energies accompanied by
large incoherent structures at higher energies. The onset of quasiparticle
damping in the spectra can be clearly traced back to the imaginary part of the
normal selfenergy $\mathrm{Im}\,\Sigma_{\vc k}(E)$ presented in
Fig.~\ref{fig:AkEcuts}(b). It shows a significant momentum dependence at low
energies, making $\mathrm{Im}\,\Sigma$ distinct below approximately 
$0.1-0.2\:\mathrm{eV}$ at the individual $\vc k$-points, however, the overall 
profiles of the selfenergy including higher energies are quite uniform 
in the Brillouin zone, with a slight modulation in the magnitude.

In the following analysis of the scattering processes, we will use the
selfenergy $\mathrm{Im}\,\Sigma$ as our primary tool. Its use is motivated by
the fact that it is directly related to the scattering rate and
semi-additively reflects the various scattering processes.
More precisely, considering for example Eq.~\eqref{eq:selfEn2}, we can
interpret the right-hand side as a straightforward sum of the scattering
contributions. This enables to resolve the scattering rate according to the
excitation involved, its characteristic wave vector, to judge the relative
importance of the interaction channels etc. by simply restricting the sum in
Eq.~\eqref{eq:selfEn2} accordingly. Practically, we first iterate the
selfenergy and Dyson's equations to arrive at a fully selfconsistent solution
and then perform the resolution by evaluating the right-hand side of
Eq.~\eqref{eq:selfEn2} with the above restriction. 
Through the selfconsistent spectral functions entering Eq.~\eqref{eq:selfEn2},
the contributions of various scattering processes get partially mixed, 
hence we call $\mathrm{Im}\,\Sigma$ {\em semi}-additive only.

\begin{figure}[tb]
\includegraphics[scale=1.0]{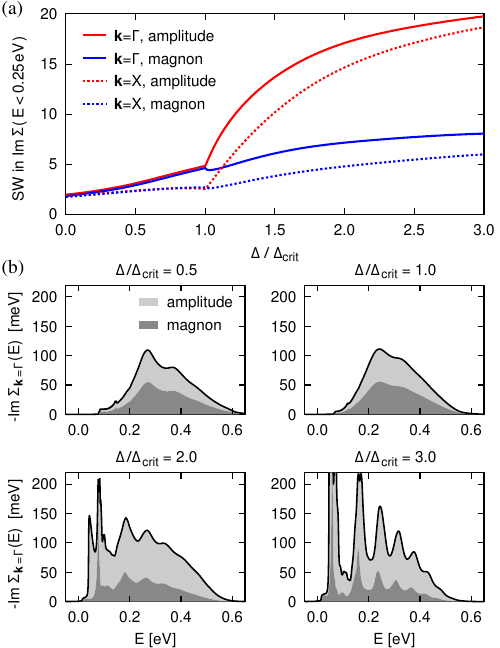}
\caption{{\bf Relative impact of the amplitude and magnon modes:}
(a)~Spectral weight of the contribution of the amplitude and magnon modes to
the imaginary part of the selfenergy. 
The spectral weight is defined as
$\mathrm{SW}=\int_0^{E_{\mathrm{max}}}
|\mathrm{Im}\,\Sigma_{\vc k}(E)|\,\mathrm{d} E$ with the cutoff
$E_\mathrm{max}=0.25\:\mathrm{eV}$
and is plotted through the phase diagram of Fig.~\ref{fig:d4excit}(a) for 
$\vc k=\Gamma$ and $\vc k=X$. A similar picture --- showing a steep onset of the
amplitude mode contribution upon entering the AF phase --- is obtained also for
other $\vc k$-points and other cutoffs $E_\mathrm{max}$.
(b)~Decomposition of the selfenergy into amplitude-mode and magnon
contributions as performed in Fig.~\ref{fig:AkEcuts}(c) but
presented here for the remaining parameter points used in 
Fig.~\ref{fig:AkEoverview}(a).
}\label{fig:amplmag}
\end{figure}

As a first example of such an analysis, Fig.~\ref{fig:AkEcuts}(c) presents
$\mathrm{Im}\,\Sigma$ for $\vc k=\Gamma$ resolved according to the
contributing mode. $\mathrm{Im}\,\Sigma$ profiles due to both amplitude mode
and magnon have a similar shape of a broad hump, with the amplitude mode
contributing about two times more than the magnon. A clear difference is
observed at low energies $\lesssim 0.1\:\mathrm{eV}$, where the contribution
due to the amplitude mode fully takes over. To explore the relative
contribution of the modes further, in Fig.~\ref{fig:amplmag}(a) we scan
through the phase diagram by showing the $\Delta$-dependent spectral
weight of $\mathrm{Im}\,\Sigma$. This integral quantity, defined as 
$\mathrm{SW}=\int_0^{E_\mathrm{max}} 
\,|\mathrm{Im}\,\Sigma_{\vc k}(E)|\,\mathrm{d}E$, reveals a strong onset
of the scattering upon entering the AF phase.
In the NM phase up to the critical point, the two magnetic excitations are
nearly equivalent (up to the small difference due to in-plane anisotropy
$\Delta'$) and they give almost identical contribution to
$\mathrm{Im}\,\Sigma$. With the emergence of the AF order for $\Delta$ above
$\Delta_\mathrm{crit}$, the amplitude and magnon modes are formed and are
found to operate differently, with the strong onset of scattering being
attributed to the amplitude mode corresponding to fluctuations of the length
of the ordered moments.  On the other hand, the magnon mode, corresponding to
the rotations of the ordered moments, continues to contribute with the same
trend as observed in the NM phase.
For the illustration, we have chosen $\vc k=\Gamma$ and $\vc k=X$ and the
energy cutoff of $E_\mathrm{max}=0.25\:\mathrm{eV}$. Quantitative differences
are observed with different choices but the gross picture remains the same.
Finally, Fig.~\ref{fig:amplmag}(b) also presents the full energy profiles of
mode-resolved $\mathrm{Im}\,\Sigma$ for the remaining parameter points used in
Fig.~\ref{fig:AkEoverview}(a), ranging from relatively featureless spectra in
the NM phase up to the strongly polaronic regime at large $\Delta$, which is
characterized by a pronounced sequence of peaks in both $\mathrm{Im}\,\Sigma$
and the spectral function~$\mathcal{A}$.


\subsection{Detailed analysis of the scattering processes}
\label{sec:scatanal}

Having observed the leading role of the amplitude mode in the scattering of an
extra electron injected into the $d^4$ AF background, we now use a finer
decomposition of $\mathrm{Im}\,\Sigma$ to identify the particular physical
mechanism behind this scattering and subsequently draw an intuitive picture of
the electron propagation.

To this end we resolve $\mathrm{Im}\,\Sigma$ calculated via
Eq.~\eqref{eq:selfEn2} not only according to the participating mode as in the
previous paragraph, but additionally according to the characteristic momentum
of the mode and the employed interaction channel in
$\mathcal{H}_{d^5\textbf{--}d^4}$. The former resolution is achieved in a
rough form simply by summing over $\vc q$ vectors spanning one of the four
areas of Brillouin zone sketched in Fig.~\ref{fig:AkEoverview}(b). The areas
equally divide the Brillouin zone and are labeled according to the
high-symmetry point they contain. The latter resolution according to the
interaction channel is achieved by reducing the matrix elements $M$ and
$\bar{M}$ in Eq.~\eqref{eq:selfEn2} so that only a particular channel is
preserved.

Let us start the discussion by a brief summary of the three interaction
channels present in $\mathcal{H}_{d^5\textbf{--}d^4}$ of \eqref{eq:Hd5d4} and
the derived $\mathcal{H}^{(1)}_{d^5\textbf{--}d^4}$ of \eqref{eq:HintAF}:
\\
\textit{(i)}~The first channel associated with the interaction parameters $A$
and $\delta A$ corresponds to bond processes, where the hopping of the doped
electron $f$ induces a flip of its \mbox{pseudospin-$\frac12$} and
simultaneously an $s\leftrightarrow T$ transition in the $d^4$ background [see
Fig.~\ref{fig:modeld5d4}(a) and Eqs.~\eqref{eq:bondSxy}, \eqref{eq:Hd5d4}].
Such a process provides a linear coupling of $f$ to the magnetic excitations
that is active across the entire phase diagram.
\\
\textit{(ii)}~The second channel, associated with $B$, represents a
pseudospin-sensitive hopping of $f$ which couples to an interchange of $T_x$ and
$T_y$ in the $d^4$ background.  This process rotates the ordered magnetic
moment in the background and translates to a linear coupling between $f$ and
magnons in the AF phase.
\\
\textit{(iii)}~Finally, in the third channel associated with the interaction
parameters $C$ and $\delta C$, the corresponding Hamiltonian term merely swaps
the $d^5$ configuration (i.e. $f$ including its state) and the $d^4$
configuration (i.e.  $s$, $T_x$, or $T_y$) initially present on the bond.
While it might appear trivial, this process in fact efficiently couples the
doped electron $f$ to the amplitude mode in the AF phase and will be later
identified as an essential ingredient for understanding the propagation of
$f$.

\begin{figure}[tb]
\includegraphics[scale=1.0]{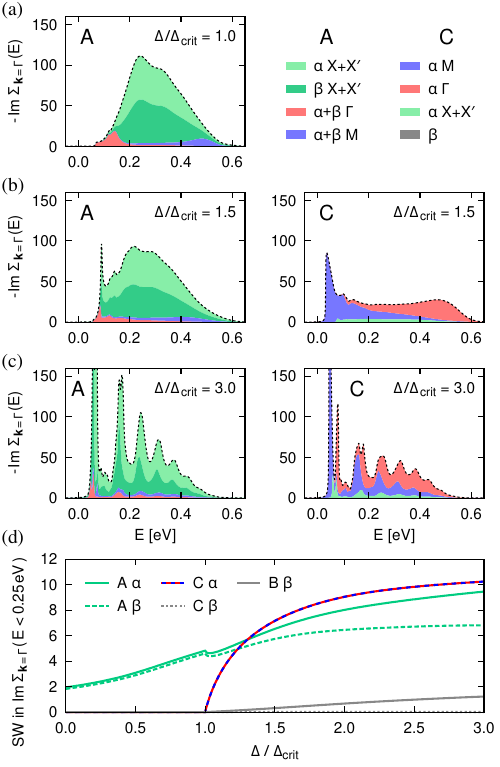}
\caption{{\bf Identification of the essential processes:}
(a)~Decomposition of the selfenergy \mbox{$\mathrm{Im}\,\Sigma_{\vc
k=\Gamma}(E)$} as obtained for $\Delta/\Delta_\mathrm{crit}=1$ into
contributions of the two modes resolved according to mode momenta and the
Hamiltonian term involved.  Mode-momentum resolution is realized by
restricting the $\vc q$ sum in the selfenergy equation \eqref{eq:selfEn2} to a
particular area in the Brillouin zone following the Brillouin-zone
partitioning shown in Fig.~\ref{fig:AkEoverview}(b).  The key on the right
indicates the selected combinations of modes and Brillouin-zone areas.  The
key is given separately for the $A$ and $C$ processes.  At
$\Delta\leq\Delta_\mathrm{crit}$, only the $A$ terms of
$\mathcal{H}_{d^5\text{--}d^4}$ are active and most of the $\vc k=\Gamma$
selfenergy is equally contributed by $\alpha$ and $\beta$ modes from the $X$
and $X'$ areas.  (b),(c)~The same for $\Delta/\Delta_\mathrm{crit}=1.5$ and
$3.0$, where the $C$ terms contribute as well and do so via the $\alpha$ mode.
(d)~$\Delta$-dependent low-energy spectral weight of 
\mbox{$\mathrm{Im}\,\Sigma_{\vc k=\Gamma}(E)$}
resolved according to the Hamiltonian term in $\mathcal{H}_{d^5\text{--}d^4}$ 
and contributing mode.
The spectral weight is defined the same way as in Fig.~\ref{fig:amplmag}(a)
with an identical cutoff.
}\label{fig:ImEdecomp}
\end{figure}

A representative set of resolved $\mathrm{Im}\,\Sigma$ spectra is shown in
Fig.~\ref{fig:ImEdecomp}, starting with the very endpoint of the NM phase
[$\Delta/\Delta_\mathrm{crit}=1$ in Fig.~\ref{fig:ImEdecomp}(a)]. Here only
the $A$ channel contributes, since the coupling between $f$ and magnetic
excitations provided by $B$, $C$ channels goes beyond the linear order
included in $\mathcal{H}^{(1)}_{d^5\textbf{--}d^4}$ of \eqref{eq:HintAF}. As
noted earlier, the contributions of the two magnetic modes is nearly equal in
the NM phase. The $\vc k=\Gamma$ point selfenergy is generated mainly by
excitations with momenta $\vc q$ around $X$ and $X'$ and a similar selectivity
is found for the other $\vc k$ points as well.  In the AF phase
[Figs.~\ref{fig:ImEdecomp}(b) and (c)], the $C$ processes start to play an
important role. Here only the amplitude mode contributes, again with a certain
selectivity in momentum space.
We can therefore conclude that the scattering of the doped electron $f$ has
two significant components: \textit{(i)} scattering via pseudospin-flip $A$
processes with a balanced contribution of both kinds of magnetic excitations,
which acts both in NM and AF phase, and \textit{(ii)} pseudospin-conserving
scattering via $C$ processes that is limited to the amplitude mode and AF
phase.
This view is further supported by the $\Delta$-dependence of the low-energy
spectral weight of $\mathrm{Im}\,\Sigma$ presented in
Fig.~\ref{fig:ImEdecomp}(d). Resolved according to the contributing process
and mode, it shows a monotonous trend for the $A$ processes and---upon
entering the AF phase---a steep onset of the amplitude mode contributing via
the $C$ channel. Contributing via both channels, the amplitude mode thus takes
over in the AF phase, as observed earlier. In Fig.~\ref{fig:ImEdecomp}(d), we
also show the relatively marginal contribution of AF magnons via the $B$
channel.

\begin{figure}[t!b]
\includegraphics[scale=1.0]{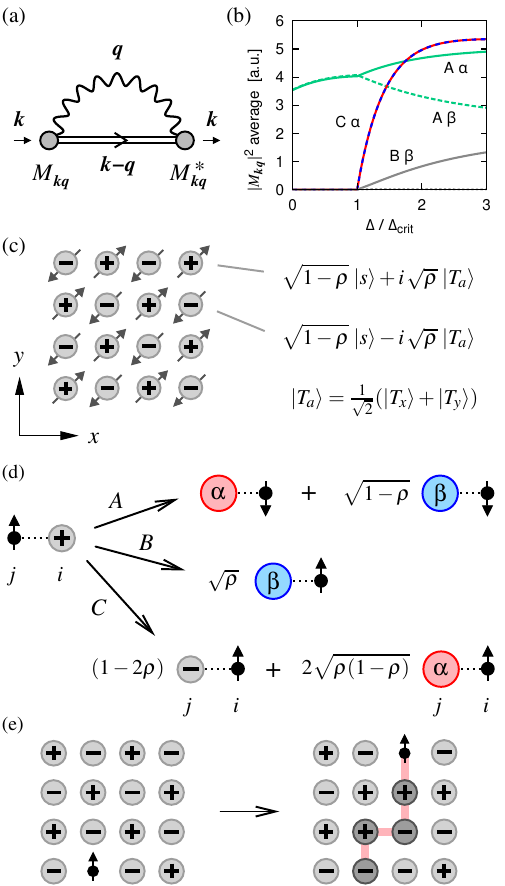}
\caption{
(a)~Momentum flow in a generic selfenergy diagram and matrix element notation.
(b)~Matrix elements $|M_{\vc k\vc q}|^2$ averaged over $\vc k$ and $\vc q$ momenta 
and resolved according to the process and mode involved.
(c)~Schematic picture of the AF condensate. Each site hosts a
superposition of singlet ($|s\rangle$) and a~particular combination of triplets 
($|T_a\rangle$) giving rise to a van Vleck moment.
(d)~Scheme summarizing the significant final states
when $\mathcal{H}_{d^5\text{--}d^4}$ of Eq.~\eqref{eq:Hd5d4} moves an electron
to a site occupied initially by the local AF ground-state superposition of
panel (c). The three propagation channels $A$, $B$, $C$ generate combinations 
of local amplitude ($\alpha$) and magnon ($\beta$) excitations with various 
relative amplitudes. A detailed description is given in the text.
(e)~Simplified picture of electron propagation via dominant $C$ processes.
During its motion, the electron constantly exchanges its position with the
local superpositions of panel (c), making them misplaced and leaving a string
of amplitude excitations behind.
}\label{fig:interp}
\end{figure}

Figure~\ref{fig:interp}(b) demonstrates that the above trends are in fact
embedded in the matrix elements $M$, $\bar{M}$ and are therefore related to
the very physical nature of the corresponding processes. Even though the
effects linked to e.g. densities of states are naturally present, they are not
essential to bring up an intuitive picture of the propagation.
The data shown in Fig.~\ref{fig:interp}(b) represent matrix elements that
result from momentum averaging motivated by the generic selfenergy diagram
depicted in Fig.~\ref{fig:interp}(a). To obtain a representative number for
the squared matrix element $|M_{\vc k\vc q}|^2$, we let $\vc k$ (momentum of
the electron), $\vc q$ (momentum of the magnetic excitation in the
intermediate state) and consequently $\vc k-\vc q$ (momentum of the electron
in the intermediate state) vary to cover the entire Brillouin zone. The
average $\sum_{\vc k \vc q} |M_{\vc k\vc q}|^2$ with $M_{\vc k\vc q}$ or
$\bar{M}_{\vc k\vc q}$ reduced to select the channel of interest is then
plotted in Fig.~\ref{fig:interp}(b) which shows a remarkable similarity to
Fig.~\ref{fig:ImEdecomp}(d).  We note, that the matrix elements $M_{\vc k\vc
q}$, $\bar{M}_{\vc k\vc q}$ are also responsible for the momentum-space
selectivity observed in Fig.~\ref{fig:ImEdecomp}. A closer inspection reveals
that the $A$ terms prefer $\vc k-\vc q$ around $X$, $X'$ while the $C$ terms
give the strongest contribution for $\vc k-\vc q$ around $\Gamma$ and $M$.
This explains the prominent momentum selectivity of the $A$ and $C$ scattering
displayed by $\mathrm{Im}\,\Sigma$ in Figs.~\ref{fig:ImEdecomp}(a)--(c).


\subsection{Intuitive picture of the polaronic motion}
\label{sec:intuitive}

In the end of this section we construct a simple cartoon picture of the
electron propagation in the AF-ordered $d^4$ background. The essential
ingredient is the picture of the ordered state based on the variational wave
function \eqref{eq:d4trial}. It misses the contribution of quantum
fluctuations, but properly captures the internal structure of the condensate of
$T$. The AF-ordered case is schematically depicted in
Fig.~\ref{fig:interp}(c). Each site carries a superposition
$\sqrt{1-\rho}\; |s\rangle \pm i\sqrt{\rho}\; |T_a\rangle$
of the singlet $s$ and the condensed combination $T_a$ of $T_x$ and $T_y$
[see Eqs. \eqref{eq:d4trial}--\eqref{eq:bosrot}].
In our $\phi=\pi/4$ case $|T_a\rangle=(|T_x\rangle+|T_y\rangle)/\sqrt2$. 
This combination is also connected to the amplitude mode $\alpha$, which may
be understood as an oscillation of the balance between $s$ and $T_a$, hence of
the condensate density $\rho$. The orthogonal combination 
$|T_b\rangle=(|T_y\rangle-|T_x\rangle)/\sqrt2$
is then connected to the magnon excitation $\beta$.
To track the electron propagation, we now consider a
bond with one site occupied by $f$ and the other site by the $d^4$
ground-state superposition, formally 
$f^\dagger_{\uparrow j} (\sqrt{1-\rho}\;s^\dagger+i\sqrt{\rho}\;T_a^\dagger)^\pd_i
\,|\mathrm{vac}\rangle
=f^\dagger_{\uparrow j} c^\dagger_i \,|\mathrm{vac}\rangle$.
By applying $\mathcal{H}_{d^5\text{--}d^4}$ \eqref{eq:Hd5d4} to this initial
configuration, we move $f$ to the other site. The analysis of the resulting
bond state will then allow us to identify the excitations created in the individual 
interaction channels. The situation is sketched in Fig.~\ref{fig:interp}(d).

The outcome in the $A$ interaction channel is a flipped pseudospin of $f$ and
a combination of amplitude and magnon excitations at the $d^4$ site.  The
contribution associated with the dominant parameter $\delta A$ [see
Fig.~\ref{fig:modeld5d4}(b)] reads explicitly as
\begin{equation}\label{eq:Aresult}
\pm \delta A\; 
\Bigl( \tfrac{1+i}{\sqrt2}\; a^\dagger + 
  \tfrac{1-i}{\sqrt2}\; \sqrt{1-\rho}\; b^\dagger \Bigr)_j
\,f^\dagger_{\downarrow i}
\,|\mathrm{vac}\rangle
\end{equation}
and a similar one is found for the minor parameter $A$.  The $\pm$ sign refers
to $x$ or $y$ bond, respectively.  Via $a^\dagger$ or $b^\dagger$, the hopping
of $f$ thus got linked to the excitation/absorption of either amplitude mode
$\alpha$ (carried by boson $a$) or magnon $\beta$ (carried by boson $b$).  By
comparing the absolute values of the prefactors, this simple analysis suggests
that the $\alpha$ and $\beta$ modes are nearly equally active, both in NM and
AF phases, consistently with the data of Fig.~\ref{fig:ImEdecomp}. In the AF
phase, the magnon contribution is a bit suppressed due to the $\sqrt{1-\rho}$
factor in \eqref{eq:Aresult}, as indeed observed in
Fig.~\ref{fig:ImEdecomp}(d).

The action of the $B$ term is easily understood by casting the corresponding
part of $\mathcal{H}_{d^5\text{--}d^4}$ into the form
$\sqrt{\rho}\,(b-b^\dagger)_j\,(f^\dagger_{\uparrow i} f_{\uparrow j}
-f^\dagger_{\downarrow i} f_{\downarrow j})$. 
The pseudospin-conserving motion of $f$ via the $B$
channel is thus directly linked to the creation/annihilation of 
magnon excitations and is limited to the AF phase with $\rho>0$.

Surprisingly, the most remarkable outcome of the $f$ hopping arises via the
$C$ interaction channel, which seems somewhat trivial in Eq.~\eqref{eq:Hd5d4}.
In analogy with Eq.~\eqref{eq:Aresult}, as the terms associated to the main
parameters $C_0$, $C$ we obtain
\begin{equation}\label{eq:Cresult}
\Bigl\{
[-C_0(1-\rho)+C\rho]\, c^\dagger 
-i (C_0+C)\sqrt{\rho(1-\rho)}\, a^\dagger 
\Bigr\}_j \,
f^\dagger_{\downarrow i}\,
|\mathrm{vac}\rangle \,.
\end{equation}
This result has two important consequences:

The first term in \eqref{eq:Cresult} perfectly fits the condensate structure
and shows that the pseudospin-conserving hopping may proceed without
disturbing the magnetic background at all, explaining the observation of a
``free'' motion even rather deep in the AF phase.  This feature stays in
contrast with the well-known motion of a carrier in the 2D $S=\frac12$
Heisenberg model~\cite{Kan89, Mar91}---though it is reminiscent of the motion
of a hole in the bilayer $S=\frac12$ antiferromagnet~\cite{Voj99, Rad12a,
Rad12b, Nyh22, Nyh23}, see also discussion in the Conclusions section.
Assuming $C_0\approx C$ we can estimate, that the amplitude of this ``free''
motion is suppressed by $(1-2\rho)$, as indicated in Fig.~\ref{fig:interp}(d). 

The second term corresponds to an excitation of an amplitude mode and,
assuming again $C_0\approx C$, it comes with a large amplitude
$2C\sqrt{\rho(1-\rho)}$. This term is therefore responsible for the large
onset of the scattering via the amplitude mode upon entering the AF phase, as
observed earlier.  In addition to \eqref{eq:Cresult}, there is also a
contribution linked to a magnon excitation:
$\pm i\, \delta C \, \sqrt{\rho}\; 
b^\dagger_j \,f^\dagger_{\downarrow i}\,|\mathrm{vac}\rangle$.
However, it comes with small prefactor $\delta C$ and is further reduced by
$\sqrt{\rho}$, so that it can be safely ignored.

Another---less precise, but easier---way to look at the action of the $C$
interaction is sketched in Fig.~\ref{fig:interp}(e).  As there is only a
little distinction between $C_x$ and $C_y$ in Eq.~\eqref{eq:Hd5d4}, one can
imagine, that the $f$ hopping in the $C$ channel simply ``moves around'' the
$T$ objects without altering the particular combination of $T_x$ and $T_y$.
Furthermore, since $C_0\approx C$, the amplitude of $s$ motion approximately
matches that of $T$.
Taken together, these two observations imply that the $C$-channel hopping of
$f$ also just ``moves around'' superpositions of $s$ and $T$ as rigid objects.
As a result, while $f$ is moving via $C$ channel, the ground-state local
combinations $\sqrt{1-\rho}\; |s\rangle \pm i\sqrt{\rho}\; |T_a\rangle$ are
preserved, but get placed in wrong position in the magnetic structure as shown
in Fig.~\ref{fig:interp}(e).
A moving electron thus leaves a string of excitations behind, like in the case
of a regular spin polaron in doped Heisenberg AF.  In contrast to the latter
case, however, the string of excitations is not composed of magnons (this
would require the involvement of $T_b$, which are not available in the $d^4$
ground state) but almost exclusively of condensate density / spin-length
fluctuations.
At first sight this simplified picture misses the fact that there is also
disruption-free motion of the doped electron, albeit with the amplitude
reduced by $(1-2\rho)$. This a bit hidden feature appears due to the overlap
between the two possible ground-state local configurations: $\sqrt{1-\rho}\;
|s\rangle \pm i\sqrt{\rho}\; |T_a\rangle$.



\section{Understanding the pseudogap:\\ The hedgehog toy-model}
\label{sec:toy}

One of the most interesting features of the obtained spectra is the fact that
for moderate values of the crystal field $\Delta$ the quasiparticle picture
seems to partially collapse and an onset of the pseudogap is well-visible at
particular momenta of the Brillouin zone -- see e.g. suppressed spectral
weight at $\sim (\pi/2, \pi/2)$ momentum in the otherwise more-or-less free
dispersion spectrum found at $\Delta = \Delta_\mathrm{crit}$ shown in
Fig.~\ref{fig:AkEoverview}(a).

Clearly, this phenomenon is caused by the interplay of the free motion
(dominant in for ``small'' values of $\Delta$) and the polaronic string-like
motion (dominant for ``large'' values of $\Delta$). However, the point is to
thoroughly understand the physical origin of this nontrivial problem. In
this section we show that one can intuitively understand the onset of a
pseudogap using a simple and exactly solvable toy model. We call this model a
``hedgehog'', for this reflects the graphical representation of the model
shown in Fig.~\ref{fig:toymodel}(a).

\subsection{Toy-model Hamiltonian and propagator}

\begin{figure*}[tb]
\includegraphics[width=18cm]{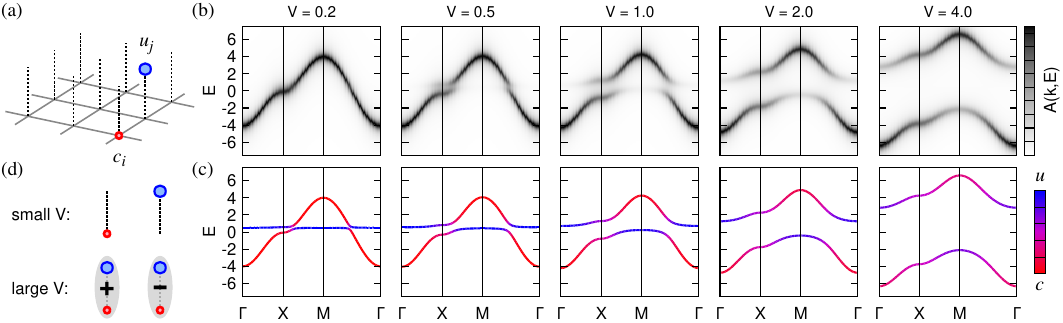}
\caption{{\bf Spectral functions of the toy model:}
(a)~Geometry of the hedgehog model---mobile fermions $c$ move on the square lattice
(bottom solid lines) via nearest neighbor hopping and hybridize with fully localized
fermions $u$ having a finite energy cost $E_u$ (the connection is suggested by dotted
lines).
(b)~Spectral function determined from the single-particle propagator
\eqref{eq:greentoymodel} of the hedgehog toy model calculated for a sequence of
increasing values of the hybridization parameter $V$. 
Apart from varying $V$, the other parameters of the toy-model are fixed to
$E_u=0.5$ and $t=1$. Large broadening $\delta=0.5$ was used to better visualize 
the spectra.
(c)~Eigenmode dispersion curves corresponding to the maps in (b).
The character of the eigenmodes is indicated by color interpolating from
red (pure $c$) to blue (purely local $u$). Even at the smallest $V=0.2$, 
there is an avoided crossing of the eigenmode dispersion (in fact, this 
crossing exists for any $V\neq 0$).
(d)~Schematic view of the eigenmodes in two limits. For a small hybridization
parameter $V$, the eigenmodes are carried by either $c$ or $u$ with a
negligible mixing between the two, apart from the area near the dispersion
crossing points. At large $V$, the eigenmodes correspond to
bonding/antibonding combinations of $c$ and $u$, weakly dispersing due to $t$.
}\label{fig:toymodel}
\end{figure*}

The Hamiltonian of the hedgehog model written below is inspired by a model 
that was introduced in \cite{Woh08}---see Eq. (4.1) {\it ibidem}. 
Following Fig.~\ref{fig:toymodel}(a), we adapt that model to the square lattice
and consider the following Hamiltonian:
\begin{equation}\label{eq:toymodel}
\mathcal{H}= 
- t \sum_{\langle ij\rangle} (c^\dagger_i c^\pd_j +\mathrm{H.c.})
+ E_u \sum_i u^\dagger_i u^\pd_i
- V \sum_i (c^\dagger_i u^\pd_i  + \mathrm{H.c.}) \,, 
\end{equation}
where the first term $\propto t$ describes hopping of spinless fermions $c$ 
along the bonds of a square lattice, the second $\propto E_u$ is the energy 
cost of creating an immobile local excitation (represented by particle $u$) 
on the ``spines'' of the hedgehog, and the last one is the hybridization 
$\propto V$ between the ``spines'' and the body of the hedgehog. 
Note that this model corresponds to the noninteracting limit of the periodic
Anderson model and can also be called a Kondo-necklace model~\cite{Bru06}.

We are solely interested in the single-particle sector of the 
Hilbert space of the Hamiltonian, which is spanned by the following basis:
\begin{align}
| \psi^{(1)}_{\vc k} \rangle = c^\dag_{\vc k} |\mathrm{vac}\rangle \,,
\quad
| \psi^{(2)}_{\vc k} \rangle = u^\dag_{\vc k} |\mathrm{vac}\rangle \,.
\end{align}
In this basis the Hamiltonian matrix takes the form:

\begin{equation} \label{eq:hmatrix}
\mathcal{H} = \sum_{\vc k}
\begin{pmatrix}
c^\dagger_{\vc k} & u^\dagger_{\vc k}
\end{pmatrix}
\begin{pmatrix}
\varepsilon_{\vc k} & -V \\
-V & E_u 
\end{pmatrix}
\begin{pmatrix}
c^\pd_{\vc k} \\ u^\pd_{\vc k}
\end{pmatrix}
\end{equation}
with the nearest-neighbor tight-binding dispersion
$\varepsilon_{\vc k}=-2t(\cos k_x+\cos k_y)$.

From the Hamiltonian in the matrix form we obtain [by inverting the
Hamiltonian matrix in Eq.~\eqref{eq:hmatrix} and taking the upper left
element] the single-particle propagator that is related to adding a fermion to
the body of the completely empty ``hedgehog'':
\begin{equation}\label{eq:greentoymodel}
G(k,E) \equiv \Big\langle \psi^{(1)}_{\vc k} \Big| 
\frac{1}{E - \mathcal{H}} \Big| \psi^{(1)}_{\vc k} \Big\rangle
= \frac{1}{E -\varepsilon_{\vc k} -\frac{V^2}{ E - E_u}}.
\end{equation}
Note that the above propagator is qualitatively equivalent to Eq.~(4.4) 
in Ref.~\cite{Woh08} ({\it modulo} the next nearest neighbor hopping and 
a different sign convention).

Before we continue, let us discuss in some more detail the connection between
the hedgehog model and the ``full'' model discussed throughout the paper. As
already noted, a similar model was postulated in \cite{Woh08}, in order to
reveal the competition between the free and string-like (polaronic) motion.
Following Ref.~\cite{Woh08}, we suggest that: \textit{(i)}~the electron motion
along the body of the hedgehog mimics the free electron motion and
\textit{(ii)}~the electron visiting spines, which costs extra energy $E_u$,
mimics the string-like motion. While point \textit{(i)} is straightforward to
grasp, one may wonder how visiting the spines may mimic the string-like
motion. As seen in Eq.~\eqref{eq:greentoymodel}, within the hedgehog model
the simple expression $V^2/(E-E_u)$ replaces the full selfenergy $\Sigma$ such
as that shown in Fig.~\ref{fig:amplmag}. In other words, the complicated
selfenergy of the full model, with broad continuum and possibly
multiple peaks, is now approximated by a single pole, leading to the most
simplistic representation of the string-like motion. Therefore, the spine
state $u$ should be interpreted here as a single state that represents the
whole spectrum of intermediate states in the complicated string-like motion
and its energy $E_u$ should be chosen as corresponding to the position of the
maximum of the selfenergy $-\mathrm{Im}\,\Sigma$ of the full model [e.g.
$E_u\approx 0.3\:\mathrm{eV}$ for $\Delta/\Delta_\mathrm{crit}=0.5$, cf.
Fig.~\ref{fig:amplmag}(b)]. Naturally, such a simple model cannot reproduce
the whole rich physics found in the full model. Nevertheless, as shown below,
it is able to distinguish between the two regimes: one dominated by the free
motion and the other by the (polaronic) selfenergy. 

\subsection{Results and discussion}

Evolution of spectral functions, as calculated from the toy-model's propagator
\eqref{eq:greentoymodel}, with increasing value of hybridization parameter $V$
is shown in Fig.~\ref{fig:toymodel}(b). We observe that the evolution of these
spectra bear a surprisingly large resemblance with the ones obtained for the
far more complex spin-orbital model [see Fig.~\ref{fig:AkEoverview}(a) above].
In particular, also here we observe three distinct regimes discussed below:

{\it Large $V$: the string limit.} Once $V\sim 1-2$ or larger we clearly observe
two separate delta-like peaks in the spectral function, both showing a reduced
bandwidth. This is the typical strong-polaron (or string-like) limit, as also
discussed in~\cite{Woh08}. In this case, irrespective of the momentum, the
two eigenstates have a string-like character: they are formed by tightly bound
$c$ and $u$ on the spines of the hedgehog [see Fig.~\ref{fig:toymodel}(d)],
largely localized and acquire a small only dispersion due to an effectively
reduced hopping $t$ along the hedgehog plane. 

{\it Moderate $V$: the pseudogap regime.} Once $V\sim 0.5-1$, we observe quite
a distinct picture: a free dispersion with a clearly visible pseudogap at
$E^\star \sim E_u$ and almost dispersionless peaks with low spectral weight
below or above $E^\star$ (cf. central middle and central bottom panels of
Fig.~\ref{fig:toymodel}). This peculiar feature is due to the fact that the
character of the eigenstates strongly depends on momentum: 
Through the Brillouin zone, the eigenstates vary their character between a
predominantly localized (for example for the lower branch this occurs at large
momenta around $M$) and an itinerant one (for the lower branch at small
momenta around $\Gamma$). 
This way the ground (excited) state minimizes (maximizes) its energy by making
either the kinetic or the string part dominant, depending on the momentum. At
the same time, as the character of the eigenstates becomes more localized,
their spectral weight contribution to the single-particle propagator decreases
(since their ``free particle character'' gets reduced).  

{\it Small $V$: the hidden pseudogap limit.} At $V\sim 0.2$ the spectrum
in Fig.~\ref{fig:toymodel}(b) seems to contain merely a free dispersion. 
However, a closer look at the underlying eigenvalues reveals that the situation
is more complex [see the left panel of
Fig.~\ref{fig:toymodel}(c)]: just as in the moderate $V$ regime there is a
pseudogap in the free dispersion. The latter arises due to the fact that again
this ``free band'' is actually formed by two eigenstates: the ground and the
excited state that switch their characters depending on momentum. As a
consequence, either the ground or the excited state shows up in the spectral
function of $c$ for a given momentum.

To sum up, the employed toy-model shows that, as a result of the interplay
between the free hopping and string-like motion, a pseudogap---or a vanishing
{\it noninteracting} quasiparticle spectral weight---may occur in the free
dispersion. Interestingly, this happens not only for moderate ratios of the
free dispersion w.r.t. the string-like motion but even once the latter type of
motion is vanishingly small. This phenomenon can also be verified by
inspecting properties of the propagator \eqref{eq:greentoymodel}: The
quasiparticle spectral weight 
$Z=1/(1- \partial \mathrm{Re} \Sigma / \partial E |_{E = E^\star} )$ 
vanishes at $E^\star=E_u$ due to 
$\partial {\rm Re} \Sigma / \partial E |_{E = E^\star} = \infty $ for
{\it any finite} value of $V$.

Altogether, the similarity between the behavior of the hedgehog toy-model,
which essentially is a noninteracting one, and the full polaronic model means
that the pseudogap-like feature may be understood within a single-band picture
as an effect of a scattering on a single energy level. Formally, this
correspondence stems from the rather crude, yet apparently viable, single-pole
approximation to the selfenergy of the full model.  


\section{Conclusions}
\label{sec:conclusions}

To conclude we have studied the motion of a single doped electron introduced
into the 2D Mott insulator with $d^4$ valence shell that is subject to
moderate to strong spin orbit coupling. In this situation, the entanglement of
spin and orbital degrees of freedom becomes essential for both the $d^5$ doped
carrier as well as for the $d^4$ local moments in the magnetic background.
We have formulated the corresponding exchange model for the undoped case by
extending earlier works, which show that the interaction between the magnetic
moments in the case of the prototypical $d^4$ spin-orbit Mott insulator
\mbox{Ca$_2$RuO$_4$} can be described in terms of a 2D XYZ model for an
effective pseudospin $\widetilde{S}=1$. The essential part of the model is
strong planar single-ion anisotropy microscopically determined by the
interplay of the spin-orbit coupling $\zeta$ and tetragonal field splitting
$\Delta$ among the $t_{2g}$ orbitals~\cite{Jai17}.  
Using $\Delta/\zeta$ as a control parameter, the model can be driven through a
quantum critical point separating its two phases---the nonmagnetic (NM) phase
at smaller $\Delta/\zeta$ and the antiferromagnetic (AF) phase for larger
$\Delta/\zeta$. The latter can be associated with a condensation of the
excitonic-like magnetic excitations hosted by the NM phase and features a
Goldstone magnon and an amplitude mode in its excitation spectrum.
When a mobile $d^5$ carrier is introduced into such a magnetic background, its
propagation proceeds via a rather complex set of processes, which include
processes associated with a direct creation/annihilation of a magnetic
excitation as well as those that apparently leave the background intact.  Here
we analyzed the competition of these two classes of processes in great detail
and uncovered their respective role in the renormalization of the doped
electron across the model phase diagram.
Experimentally, the studied model at $\Delta/\zeta \sim 1.5-2.0$, i.e.
moderately deep in the AF phase, describes the inverse photoemission performed
on \mbox{Ca$_2$RuO$_4$}.

The two main results of this work are as follows: 

{\it First}, both on the NM side of the quantum critical point as well as just
past it, i.e. for the weakly ordered AF case, the doped carrier largely
behaves as a free quasiparticle. In this regime the electron propagates via a
combination of free hopping, which does not disturb the $d^4$ background built
mostly of ionic singlets, and scattering processes originating in the
spin-spin channel, i.e. from the coupling of pseudospin-$\frac12$ carried by
the electron to the pseudospins $\widetilde{S}=1$ of the background. This
scattering employs---without a particular preference---all kinds of magnetic
excitations. It operates through the entire phase diagram, but in the above
case and within our parameter regime, it is by itself not sufficient to induce
a strong polaronic behavior of the doped carrier. Nevertheless, the coupling
to the magnetic excitations is still visible in the spectrum. Namely, at
particular values of momenta a pseudogap develops in the quasiparticle
dispersion and this can be attributed to the onset of a string-like polaronic
motion. We have understood the interplay of the latter and the free dispersion
in quite some detail using a (hedgehog) toy-model. 

{\it Second}, deeper in the AF phase the character of the spectrum
qualitatively changes and strong ladder-like features, that are signatures of
the onset of the string potential~\cite{Bul68, Kan89, Bie19}, are visible.
While this result describes the realistic case of \mbox{Ca$_2$RuO$_4$}, from
the general point of view it is interesting in itself to see such a strong
onset of the string potential in the model.
Our detailed analysis shows that the strong string potential arises here due
to the dominant coupling of the doped electron to the amplitude mode in the
density-density channel. This type of scattering starts upon entry into the AF
phase with the appearance of a magnetic condensate and steeply rises
afterwards, following the growth of the condensate amplitude.

In a broader context, the AF-phase results are somewhat reminiscent of the
observed coupling of the doped carrier to the spin fluctuations deep in the
antiferromagnetic phase of the bilayer Heisenberg magnets~\cite{Voj99, Rad12a,
Rad12b, Nyh22, Nyh23}. These are described by singlet-triplet models that
share certain features with our model, such as the magnetic condensation and
the structure of excitation spectra, though the details of the respective
models are different.
On the other hand, we note that our case is distinct from the ruthenate
photoemission problem modeled in~\cite{Klo20}. There, the coupling to the
amplitude mode was neglected, for only the magnon mode of Ref.~\cite{Oit08}
was considered. Nevertheless, the ladder spectrum and the string potential was
also observed, being traced back to the strong magnetic and hopping
anisotropies of the $t$--$J$ model used.


\acknowledgments

We would like to thank G.~Khaliullin for useful discussions. J.R. and J.Ch.
acknowledge support by Czech Science Foundation (GA\v{C}R) under Project
No.~GA22-28797S and by the project Quantum Materials for Applications in
Sustainable Technologies, Grant No.~CZ.02.01.01/00/22\_008/0004572.
Computational resources were provided by the e-INFRA CZ project (ID:90254),
supported by the Ministry of Education, Youth and Sports of the Czech
Republic. K.W. thanks IIT Madras for an IoE Visiting Scientist position which
enabled the completion of this work.


\section*{Data availability}

The data that support the findings of this article are openly available
\cite{repository}.


\appendix

\section{Microscopic formulas for $d^4$ exchange parameters}
\label{app:d4}

The exchange parameters entering the effective \mbox{pseudospin-1} model \eqref{eq:Hd4} 
that drives the $d^4$ magnetic background are given explicitly by
\begin{align}
J&=\frac{7\!-\!c_0\!+\!c_1\!-\!7c_0c_1\!-\!2\sqrt{2}s_0 s_1}{48(1-3\eta)}+ \notag \\
 &\qquad+\frac{13\!-\!c_0\!-\!5c_1\!+\!29c_0c_1\!+\!13\sqrt{2}s_0s_1}{96}+ \notag \\
 &\qquad\qquad+\frac{9\!-\!c_0\!-\!c_1\!+\!5c_0c_1\!+\!3\sqrt{2}s_0s_1}{32(1+2\eta)}
\,, \\
\rule{0mm}{10mm}
\delta J&=\frac{5\!-\!3c_0\!+\!3c_1\!-\!5c_0c_1\!+\!2\sqrt{2}s_0s_1}{48(1-3\eta)}+ \notag \\
 &\qquad+\frac{-10\!-\!6c_1\!+\!4c_0c_1\!-\!7\sqrt{2}s_0s_1}{96}- \notag \\
 &\qquad\qquad-\frac{2c_0\!+\!2c_0c_1\!+\!\sqrt{2}s_0s_1}{32(1+2\eta)}
\,, \\
\rule{0mm}{10mm}
J_z&=\frac{1-\cos 4\vartheta_1}{24(1-3\eta)} 
 +\frac{17-24c_1+7\cos 4\vartheta_1}{96}+ \notag \\
 &\qquad\qquad\qquad+\frac{7-8c_1+\cos 4\vartheta_1}{32(1+2\eta)}
\,.
\end{align}
The respective interaction constants are given in
units of $t^2/U$ and using a shorthand notation $c_{0,1}=\cos 2\vartheta_{0,1}$,
$s_{0,1}=\sin 2\vartheta_{0,1}$, and $\eta=J_\mathrm{H}/U$.
Here $U$ denotes the usual intraorbital Hubbard repulsion and $J_\mathrm{H}$
Hund's coupling.


\section{Microscopic formulas for $d^5$--$d^4$ interaction parameters}
\label{app:d5d4}

Interaction parameters of the coupling between $d^5$ electron-like carriers
and $d^4$ background used in Eq.~\eqref{eq:Hd5d4} may be compactly written
as follows:
\begin{align}
A &= \left[2c^2c_0-(cs_0+sc_0)s\right]\tfrac1{\sqrt2}c_1 \,, \\
\delta A &= (cs_0+sc_0)(cs_1+\tfrac1{\sqrt2}sc_1) \,, \\
B &= (cs_1^2-cc_1^2+\sqrt2sc_1s_1)c  \,, \\
C_0 &= c^2+c_0^2+2csc_0s_0 \,, \\
C &= 1-s^2s_1^2+\sqrt2 csc_1s_1 \,, \\
\delta C &= (cs_1+\tfrac1{\sqrt2}sc_1)\sqrt2 sc_1 \,.
\end{align}
They are given in units of $t/4$ and expressed as polynomials in the
wave-function factors
$c=\cos\vartheta$, $s=\sin\vartheta$,
$c_0=\cos\vartheta_0$, $s_0=\sin\vartheta_0$,
$c_1=\cos\vartheta_1$, and $s_1=\sin\vartheta_1$.


\section{Matrix elements of the coupling between $f$ and elementary excitations $\alpha$, $\beta$}
\label{app:d5d4fab}

The matrix elements entering the linear interaction of $\alpha$, $\beta$
elementary excitations with the doped electrons in Eq.~\eqref{eq:HintAF}
read as
\begin{align}
M_{\vc k\vc q s}^\alpha &= 
4i\,( W_{\vc k-\vc q}^{\alpha s} u_{\alpha\vc q}-W_{\vc k}^{\alpha s} v_{\alpha\vc q} ) \,, \notag \\
M_{\vc k\vc q s}^\beta &= 
4i\,( W_{\vc k-\vc q}^{\beta s} u_{\beta\vc q}-W_{\vc k}^{\beta s} v_{\beta\vc q} ) \,, \notag \\
\bar{M}_{\vc k\vc q s}^\alpha &= 
4i\,( \bar{W}_{\vc k-\vc q}^{\alpha} u_{\alpha\vc q}+\bar{W}_{\vc k}^{\alpha} v_{\alpha\vc q} ) \,, \notag \\
\bar{M}_{\vc k\vc q s}^\beta &= 
4i\,( \bar{W}_{1\vc k-\vc q}^{\beta s} u_{\beta\vc q}+\bar{W}_{2\vc k}^{\beta s} v_{\beta\vc q} ) \,.
\end{align}
They combine the Bogoliubov factors $u_{\alpha\vc q}$, $v_{\alpha\vc q}$,
$u_{\beta\vc q}$, $v_{\beta\vc q}$ [see Eq.~\eqref{eq:Bogol}] together with
\begin{align}
W^{\alpha s}_{\vc k} &=  \delta A\,\eta_{\vc k} (-\cos\phi+is\sin\phi)
                         -A\,\gamma_{\vc k}(\cos\phi+is\sin\phi) \,, \notag \\
W^{\beta s}_{\vc k} &= [\delta A\,\eta_{\vc k} (\sin\phi+is\cos\phi) 
                        +A\,\gamma_{\vc k} (\sin\phi-is\cos\phi) ]\cos\theta \,, \notag \\
\bar{W}^\alpha_{\vc k} &= [\, -(C_0+C)\,\gamma_{\vc k} + \delta C\, \eta_{\vc k} \cos 2\phi\,] \sin\theta\cos\theta \,, \notag \\
\bar{W}^{\beta s}_{1\vc k} &= ( + is B\,\gamma_{\vc k} - \delta C\,\eta_{\vc k} \sin 2\phi) \sin\theta \,, \notag \\
\bar{W}^{\beta s}_{2\vc k} &= ( - is B\,\gamma_{\vc k} - \delta C\,\eta_{\vc k} \sin 2\phi) \sin\theta \,.
\end{align}
In these formulas, $s$ has to be understood as
$s=\pm 1$ for pseudospin $\uparrow$, $\downarrow$, respectively.
The momentum dependence of the matrix elements is captured using the
$s$-wave and $d$-wave nearest-neighbor factors for the square lattice:
$\gamma_{\vc k}=\frac12(\cos k_x + \cos k_y)$
and 
$\eta_{\vc k}=\frac12(\cos k_x - \cos k_y)$.
Note that all $\bar{W}$ and consequently the $\bar{M}$ matrix elements 
corresponding to pseudospin-conserving terms in Eq.~\eqref{eq:HintAF}
contain $\sin\theta$ making them to disappear in the nonmagnetic phase
and to grow as $\sqrt{\rho}$ upon entering the AF phase.
The simplified formulas for the nonmagnetic phase and $\Delta'=0$ 
that are relevant to Eq.~\eqref{eq:HintPM} are obtained by setting 
$\theta=0$, $\phi=0$.


\bibliography{paper}

\end{document}